\documentclass[openbib,12pt]{article}

\newcommand{\utwi}[1]{\mbox{\boldmath $ #1$}}

\usepackage{booktabs,dcolumn}
\usepackage{psfrag,graphicx}
\usepackage{eso-pic,graphicx}
\usepackage{amsmath}
\usepackage{graphicx}
\usepackage{pdflscape}
\usepackage{longtable}
\usepackage{epstopdf}
\usepackage{appendix}
\usepackage{dashbox}

\newcolumntype{z}[1]{D{.}{.}{#1}}

\newcommand{\cb}{\textcolor{blue}}
\date{}

\begin{document}

\title{
\begin{center} {\Large \bf A semi-parametric realized joint value-at-risk and expected shortfall regression framework} \end{center}}
 \author{Chao Wang$^{1}$, Richard Gerlach$^{1}$, Qian Chen$^{2}$  
 \\
 $^{1}$Discipline of Business Analytics, The University of Sydney\\
 $^{2}$Business School, Shenzhen Technology University}

\date{} \maketitle

\begin{abstract}
\noindent
A new realized conditional autoregressive Value-at-Risk (VaR) framework is proposed, through incorporating a measurement equation into the original quantile regression model. The framework is further extended by employing various Expected Shortfall (ES) components, to jointly estimate and forecast VaR and ES. The measurement equation models the contemporaneous dependence between the realized measure (i.e., Realized Variance and Realized Range) and the latent conditional ES. An adaptive Bayesian Markov Chain Monte Carlo method is employed for estimation and forecasting, the properties of which are assessed and compared with maximum likelihood through a simulation study. In a comprehensive forecasting study on 1\% and 2.5 \% quantile levels, the proposed models are compared to a range of parametric, non-parametric and semi-parametric models, based on 7 market indices and 7 individual assets. One-day-ahead VaR and ES forecasting results favor the proposed models, especially when incorporating the sub-sampled Realized Variance and the sub-sampled Realized Range in the model.

\vspace{0.5cm}

\noindent {\it Keywords}: Quantile Regression, Realized Measure, Markov Chain Monte Carlo, Value-at-Risk, Expected Shortfall.
\end{abstract}

\newpage
\pagenumbering{arabic}

{\centering
\section{\normalsize INTRODUCTION}\label{introduction_sec}
\par
}
\noindent

Since their introduction by J.P. Morgan in the RiskMetrics model at 1993, Value-at-Risk (VaR) measures have been widely employed by financial institutions and corporations around the world to assist their decision making in relation to capital allocation and risk management. VaR is a quantitative tool to measure and control financial risk that represents market risk as one number and has become a standard measurement for capital allocation and risk management. Let $\mathcal{I}_t$ be the information available at time $t$ and
\[
F_{t}(r)=Pr(r_{t}\leq r | \mathcal{I}_{t-1})
\]
be the Cumulative Distribution Function (CDF) of return $r_{t}$ conditional on $\mathcal{I}_{t-1}$. We assume that $F_{t}(.)$ is strictly increasing and continuous on the real line $\Re$. Under this assumption, the one-step-ahead $\alpha$ level Value-at-Risk at time $t$ can be defined as:
\begin{equation} \label{var_def}
Q_{t}=F^{-1}_{t}(\alpha)\qquad 0 <\alpha <1. \nonumber
\end{equation}

However, VaR has been subject to criticism because it cannot measure the expected loss for violations and is not mathematically coherent, in that it can favor non-diversification. Expected Shortfall (ES), proposed by Artzner {et al.} (1997, 1999), gives the expected loss, conditional on returns exceeding a VaR threshold, and is a coherent measure; thus, in recent years it has become more widely employed for tail risk measurement and is now favored by the Basel Committee on Banking Supervision. Within the same framework as above, the one-step-ahead $\alpha$ level Expected Shortfall can be shown (see Acerbi and Tasche, 2002, among others) to be equal to the tail conditional expectation of $r_t$:
\[
ES_{t}=E(r_t|r_t\leq Q_{t}, \mathcal{I}_{t-1}).
\]



The Basel III Accord, which was implemented in 2019, places new emphasis on ES. Its recommendations for market risk management are illustrated in the 2019 document \textit{Minimum Capital Requirements for Market Risk} that says: ``ES must be computed on a daily basis for the bank-wide internal models to determine market risk capital requirements. ES must also be computed on a daily basis for each trading desk that uses the internal models approach (IMA).''; ``In calculating ES, a bank must use a 97.5th percentile, one-tailed confidence level'' (Basel Committee on Banking Supervision 2019, p. 89). Therefore, in the empirical application of our paper, we focus on one-step-ahead tail risk forecasting at the 2.5\% quantile level. To study the performance of the tail risk models, we also test the more extreme 1\% quantile level.

Volatility estimation and prediction play a key role in forming accurate VaR or ES forecasts. Since the introduction of the Auto-Regressive Conditional Heteroskedastic (ARCH) model of Engle (1982) and the generalized (G)ARCH of Bollerslev (1986), both employing squared returns as model input, many different volatility measures and models have been developed.
Parkinson (1980) and Garman and Klass (1980) propose the daily high--low range as a more efficient volatility estimator
than daily squared returns. The availability of high frequency intra-day data has generated several popular and efficient realized measures, including Realized Variance (RV) (Andersen and Bollerslev, 1998, Andersen {et al.} 2003), and Realized Range (RR) (Martens and van Dijk, 2007; Christensen and Podolskij, 2007). In order to deal with the well-known, inherent micro-structure noise accompanying high frequency volatility measures, Zhang et al. (2005) and Martens and van Dijk (2007) have each designed sub-sampling and scaling processes, aiming to provide smoother and more efficient realized measures.

Hansen et al. (2012) extend the parametric GARCH model framework by proposing the Realized-GARCH, adding a measurement
equation that contemporaneously links unobserved volatility with a realized measure. Gerlach and Wang (2016) extend the Realized-GARCH model
by employing RR as the realized measure and illustrate that the proposed Realized-GARCH-RR framework can generate
more accurate and efficient volatility, as well as VaR and ES forecasts, compared to traditional GARCH and Realized-GARCH models. Hansen and Huang (2016) have recently
extended the parametric Realized-GARCH framework to include multiple realized measures.

The tail-risk forecast performance of  parametric volatility models heavily depends on the choice of error distribution. The non-parametric conditional autoregressive VaR
(CAViaR) models proposed by Engle and Manganelli (2004) can estimate quantile (VaR) directly without a return distribution assumption. Gerlach et al. (2011) generalize CAViaR models to a fully nonlinear family under a semi-parametric framework, incorporating the asymmetric Laplace (AL) distribution for the likelihood construction. However,
CAViaR type models cannot directly estimate ES. A joint semi-parametric model that directly estimates both VaR and ES, referred to here as
the ES-CAViaR model, is proposed by Taylor (2019).  Through incorporating an AL distribution with a time-varying scale, the likelihood can be built to enable the joint estimation of the conditional VaR
and conditional ES in this framework. Fissler and Ziegel (2016) develop a family of joint loss functions (or ``scoring rules'') for the associated VaR and ES series that are strictly consistent for the true VaR and ES series, that is, they are uniquely minimized by the true VaR and ES series. Applying specific choices of functions from the join loss functions of Fissler and Ziegel (2016), it can be shown that such a loss function is exactly the same as the negative of the AL log-likelihood function presented in Taylor (2019). Patton et al. (2019) propose new dynamics models for VaR and ES through adopting the generalized autoregressive score (GAS) framework (Creal et al., 2013; Harvey, 2013) and utilizing the loss functions in Fissler and Ziegel (2016). Gerlach and Wang (2020a) extend the ES-CAViaR model by incorporating realized measures as exogenous variables in order to improve the VaR and ES forecasting accuracy.

The main contributions of the paper are as follows. Firstly, we propose a new non-parametric realized conditional autoregressive VaR framework (Realized-CAViaR). We show that the proposed Realized-CAViaR includes the Realized-GARCH as a special case, in terms of the VaR estimation and forecast. Secondly, the Realized-CAViaR framework is extended though incorporating various ES components, in order to jointly estimate and forecast VaR and ES, motivated by Hansen et al. (2012), Taylor (2019) and Gerlach and Wang (2020a). This new framework is called Realized-ES-CAViaR, which includes a measurement equation to model the dependence between the realized measure and the latent ES series. In addition, the scaled and sub-sampled realized measures are employed, to tackle the micro-structure noise and potential inefficiency. Moreover, an adaptive Bayesian MCMC algorithm is adopted to estimate the proposed models, extending that in Gerlach and Wang (2016). In the empirical study, the proposed Realized-ES-CAViaR models employing various realized measures as inputs are assessed via their VaR and ES forecasting performance. Over the forecast period 2008--2016, the empirical results illustrate that Realized-ES-CAViaR models out-perform both the existing ES-CAViaR models and a range of competing models and methods, such as the standard GARCH, realized GARCH models and conditional autoregressive Expectile (CARE) (Taylor, 2008).

The paper is organized as follows. Section \ref{model_review} briefly reviews the existing ES-CAViaR and Realized-GARCH models. Section \ref{model_section} proposes the Realized-CAViaR and Realized-ES-CAViaR class of
models. The associated likelihood and the adaptive Bayesian MCMC algorithm for parameter estimation are presented in Section \ref{beyesian_estimation_section}. Section \ref{data_empirical_section} reviews the employed realized measures and presents the forecasting study and backtesting results. Section \ref{conclusion_section} concludes the paper and discusses future work.

\vspace{0.5cm}
{\centering
\section{\normalsize ES-CAVIAR AND REALIZED-GARCH MODELS} \label{model_review}
}
\noindent

\subsection{ES-CAViaR}\label{es_caviar_section}

Koenker and Machado (1999) note that the usual quantile regression estimator is equivalent to a maximum likelihood estimator when assuming that the data are conditionally AL
with a mode at the quantile, that is, if $r_t$ is the return data on day $t$ and $Pr(r_t < Q_t | \mathcal{I}_{t-1}) = \alpha$ then the parameters in the model for $Q_t$ can be estimated using a likelihood based on:
$$ p(r_t| \mathcal{I}_{t-1}) = \frac{\alpha (1-\alpha)}{\sigma_t} \exp \left( \frac{  -(r_t-Q_t)(\alpha - I(r_t \leq Q_t)) } {\sigma_t}\right) \,\, , $$
for $t=1,\ldots,T$ and where $\sigma_t$ is the scale parameter.

Taylor (2019) extends this result to incorporate the associated ES quantity into the conditional density. By noting a link between $ES_t$ and a dynamic $\sigma_t$, the conditional density function, extended to allow heteroskedasticity, becomes:

\begin{equation}\label{al_log}
p(r_t| \mathcal{I}_{t-1}) = \frac{(\alpha-1)}{ES_t} \exp \left( \frac{(r_t-Q_t)(\alpha - I(r_t \leq  Q_t))}{\alpha ES_t}  \right) ,
\end{equation}

allowing a likelihood function to be built, given model expressions for $Q_t$ and $ES_t$ and a zero mean return. Taylor (2019) notes that the negative logarithm of the resulting
likelihood function is strictly consistent for $Q_t$ and $ES_t$ considered jointly, that is, it fits into the class of strictly consistent joint functions for VaR and ES developed
by Fissler and Zeigel (2016).

Taylor (2019) proposes two different formulations for the dynamics between VaR and ES, which also avoid ES estimates crossing the corresponding VaR estimates, as presented
in (\ref{es_caviar_ar_model}) (ES-CAViaR-Add: ES-CAViaR with an additive VaR to ES component) and (\ref{es_caviar_exp_model})
(ES-CAViaR-Mult: ES-CAViaR with a multiplicative VaR to ES relationship):

\begin{eqnarray} \label{es_caviar_ar_model}
&&Q_{t}= \beta_0+ \beta_1 |r_{t-1}| + \beta_2 Q_{t-1},\\ \nonumber
&& \text{ES}_t= Q_t-w_t, \\ \nonumber
&& w_t=
\begin{cases}
    \gamma_0 + \gamma_1 (Q_{t-1} - r_{t-1}) + \gamma_2 w_{t-1} & \text{if } r_{t-1} \leq Q_{t-1};\\
    w_{t-1}              & \text{otherwise},
\end{cases}
\end{eqnarray}
where $\gamma_0 \ge 0, \gamma_1 \ge 0, \gamma_2 \ge 0$ to ensure that the VaR and ES estimates do not cross.

\begin{eqnarray} \label{es_caviar_exp_model}
&&Q_{t}= \beta_0+ \beta_1 |r_{t-1}| + \beta_2 Q_{t-1},\\ \nonumber
&& \text{ES}_t= (1+\exp(\gamma_0)) Q_t, \\ \nonumber
\end{eqnarray}
where $\gamma_0$ is unconstrained.





\subsection{Realized-GARCH}\label{re_garch_section}

The Realized-GARCH framework is proposed in Hansen et al. (2012), who employ realized volatility and the realized kernel as realized measures in their model. Compared to the conventional GARCH model, the Realized-GARCH incorporates a measurement equation, which captures the contemporaneous relation between unobserved volatility and a
realized measure. Studies by, for example, Hansen et al. (2012), Watanabe (2012) and Gerlach and Wang (2016) demonstrate the superiority of the Realized-GARCH compared to GARCH and GARCH-X. The main advantage of a measurement equation is that more information about the latent volatility can be incorporated into the likelihood. Furthermore, asymmetric effects of positive and negative return shocks on the volatility are incorporated. The absolute value Realized-GARCH (Abs-Realized-GARCH) specification can be written as:

\noindent
\textbf{Abs-Realized-GARCH}
\begin{eqnarray}\label{rgarch}
&& r_t= \sigma_t z_t \, , \\ \nonumber
&& \sigma_t= \beta_0 + \beta_1 X_{t-1} + \beta_2 \sigma_{t-1} \, , \\ \nonumber
&& X_t = \xi +\varphi \sigma_t + \tau_1 z_t + \tau_2 (z_t^2-1)+  u_t \, , \nonumber
\end{eqnarray}

where $X_t$ is a realized measure on volatility $\sigma_t$ scale, that is, the square root of realized variance. The third equation is called the \emph{measurement equation}. Here $z_t \stackrel{\rm i.i.d.} {\sim} D(0,1)$ and $u_t \stackrel{\rm i.i.d.} {\sim} D(0,\sigma_{u}^2)$ and $D$ is chosen to be Gaussian distribution in Hansen et al. (2012).

\vspace{0.5cm}
{\centering
\section{\normalsize MODEL PROPOSED} \label{model_section}
}
\noindent

\subsection{Realized-CAViaR Models}\label{realized_cav_model_section}

In the Abs-Realized-GARCH framework (\ref{rgarch}) with Gaussian return error distribution, we have $Q_{t}= \Phi^{-1}(\alpha) \sigma_t$, where $\Phi^{-1}(\alpha)$ is the
standard Gaussian inverse cdf. Therefore, to generalize the Abs-Realized-GARCH framework with any constant conditional return distributions, we have $Q_t = a_{\alpha} \sigma_t$, where $a_{\alpha}$ is constant scaling factor equal to the inverse CDF of given distribution. Substituting
$\sigma_t= \frac{Q_t} {a_{\alpha}}$ into the Abs-Realized-GARCH framework (\ref{rgarch}) and removing the return equation, the non-parametric Realized-CAViaR framework is proposed as:

\textbf{Realized-CAViaR:}
\begin{eqnarray}\label{re_caviar_model}
&& Q_t= \beta_0 + \beta_1 X_{t-1} + \beta_2 Q_{t-1} \, , \\ \nonumber
&& X_t = \xi +\varphi Q_t +  \tau_1 \epsilon_t + \tau_2 (\epsilon_t^2-E(\epsilon^2))+  u_t \, . \nonumber
\end{eqnarray}

In terms of the VaR estimation and forecast, the Realized-CAViaR includes the Abs-Realized-GARCH framework (\ref{rgarch}) as a special case with a non-parametric manner, by noting there is no parametric distribution assumption on the return. The multiplicative error $\epsilon_t= \frac{r_t} {Q_t}$ in the measurement equation is employed to capture the well known leverage effect, which is discussed in detail in the next section.

\subsection{Realized-ES-CAViaR Models}

The proposed Realized-CAViaR framework (\ref{re_caviar_model}) cannot directly estimate and forecast ES, hence in this section we further extend the Realized-CAViaR to a joint realized VaR and ES regression framework that can estimate and forecast vaR and ES simultaneously.

Given any constant conditional return distributions with zero mean:
\begin{eqnarray*}
Q_t &=& a_{\alpha}\sigma_t \,\,;\,\, ES_t = b_{\alpha}\sigma_t, \\
\frac{ES_t}{Q_t} &=&  \frac{b_{\alpha}}{a_{\alpha}} = c_{\alpha} \,\, \text{(i.e., constant)},
\end{eqnarray*}
where $a_{\alpha}$, $b_{\alpha}$ and $c_{\alpha}\ge 1$ are constant scaling factors dependent on the distribution.

In the ES-CAViaR-Mult framework (\ref{es_caviar_exp_model}), Taylor (2019) uses $c_{\alpha}= (1+\exp(\gamma_0))$. Following such choice of $c_{\alpha}$, we further extend the Realized-CAViaR framework to the Realized-ES-CAViaR-Mult, which can jointly estimate and forecast VaR and ES. The measurement equation is updated through linking the latent ES with the realized measures:

\noindent \textbf{Realized-ES-CAViaR-Mult:}

\begin{eqnarray} \label{re_es_caviar_exp}
&&Q_{t}= \beta_0 + \beta_1 X_{t-1} + \beta_2 Q_{t-1},\\ \nonumber
&& \text{ES}_t= (1+\exp(\gamma_0)) Q_t, \\ \nonumber
&&X_t= \xi+\phi |\text{ES}_t|+ \tau_1 \epsilon_t + \tau_2 (\epsilon_t^2-E(\epsilon^2)) + u_t. \,
\end{eqnarray}

In addition, following the ES-CAViaR-Add framework (\ref{es_caviar_ar_model}), the Realized-ES-CAViaR-Add framework is proposed as:

\noindent \textbf{Realized-ES-CAViaR-Add:}

\begin{eqnarray} \label{re_es_caviar_ar}
&& Q_{t}= \beta_0 + \beta_1 X_{t-1} + \beta_2 Q_{t-1}, \\ \nonumber
&& w_t=
\begin{cases}
    \gamma_0 + \gamma_1 (Q_{t-1} - r_{t-1}) + \gamma_2 w_{t-1} , &  r_{t-1} \leq Q_{t-1}; \\
    w_{t-1},              & \text{otherwise},
\end{cases}\\ \nonumber
&& \text{ES}_t= Q_t - w_t, \\ \nonumber
&& X_t= \xi+\phi |\text{ES}_t|+ \tau_1 \epsilon_t + \tau_2 (\epsilon_t^2-E(\epsilon^2)) + u_t, \,
\end{eqnarray}

where $X_t$ represents a realized measure on volatility scale. Besides the ES components in the models, the equations in (\ref{re_es_caviar_exp}) and (\ref{re_es_caviar_ar}) are: the
\emph{quantile equation} (for $Q_t$) and the \emph{measurement equation} (for $X_t$). Contino and Gerlach (2017) consider different distributions, that is, Student-t, for the measurement equation error $u_t$. They find that changing this distribution has no real impact on the performance of the model. Therefore, in our paper we use $u_t \stackrel{\rm i.i.d.} {\sim} N(0,\sigma_{u}^2)$.

The multiplicative error $\epsilon_t= \frac{r_t} {Q_t}$ in the measurement equation is employed to capture the well known leverage effect. As discussed in Section \ref{realized_cav_model_section}, in a parametric setting we have $Q_t = a_{\alpha} \sigma_t$, where $a_{\alpha}$ is constant scaling factor equals to the inverse CDF of given return distribution $z_t$. Therefore,
$$E(\epsilon_t) = E\left(\frac{r_t} {Q_t}\right) = E\left( \frac{r_t} {a_{\alpha} \sigma_t}\right) = E\left( \frac{\sigma_t z_t} {a_{\alpha} \sigma_t} \right)= E\left(\frac{z_t} {a_{\alpha}}\right)=\frac{E(z_t)} {a_{\alpha}}= 0.$$

In addition, to keep a zero mean asymmetry term, $(\epsilon_t^2-E(\epsilon^2) )$, we need to know $$ E(\epsilon^2) = E\left(\frac{r_t^2}{Q_t^2}\right). $$

The Realized-ES-CAViaR models do not include this second moment information. Instead, an empirical estimate is proposed, via
$E(\epsilon^2) \approx \bar{\epsilon^2}$, being the sample mean of the squared multiplicative errors, so that $E(\epsilon_t^2-\bar{\epsilon^2})= 0$ is preserved
(since $\bar{\epsilon^2}$ is unbiased). Thus, the term $\tau_1 \epsilon_t + \tau_2 (\epsilon_t^2-\bar{\epsilon^2})$ still generates an asymmetric response in
the quantile series to return shocks. Further, the sign of $\tau_1$ is expected to be opposite to that from a Realized-GARCH model, since the quantile $Q_t$ is negative
for the lower quantile levels , for example, $\alpha=1\%; 2.5\%$, considered in the paper.

Lastly, as in Taylor (2019), the resulting likelihood expression, built from:
$$ p(r_t| \mathcal{I}_{t-1}) = \frac{(\alpha-1)}{ES_t} \exp \left( \frac{(r_t-Q_t)(\alpha - I(r_t \leq  Q_t))}{\alpha ES_t}  \right)\,\, , $$
implies that $r_t - Q_t$ follows an AL distribution with time varying scale. However, we would like to emphasize that the framework does not rely on an AL or any distribution assumption
for the returns. It is simply the assumption that leads to a psuedo-likelihood, optimized at the same parameter values that minimize the associated joint VaR and ES loss function.

\subsection{Realized-ES-X-CAViaR-X Models}

As discussed in Gerlach and Wang (2020a), the two specifications above restrict the manner in which the realized measure influences the ES, that is, $X_t$ only directly appears in the quantile regression equation. Therefore, in this paper we further extend the Realized-ES-CAViaR by introducing an additive model whereby the realized measure can influence and affect the VaR and ES separately, via re-specifying the dynamics on $w_t = Q_t - ES_t$ to be directly driven by the realized measure, following Gerlach and Wang (2020a).

Still, for any zero mean constant conditional return distributions:
\begin{eqnarray*}
Q_t &=& a_{\alpha}\sigma_t \,\,;\,\, ES_t = b_{\alpha}\sigma_t \\
w_t &=& Q_t - ES_t = (a_{\alpha}-b_{\alpha})\sigma_t
\end{eqnarray*}
where $a_{\alpha}$ and $b_{\alpha}$ are constant scaling factors dependent on the distribution.

\bigskip
Then, for an absolute value GARCH-X model:
\[
\sigma_t = \beta_0 + \beta_1 X_{t-1} + \beta_2 \sigma_{t-1}.
\]

It follows that:
\begin{eqnarray} \nonumber
w_t &=& (a_{\alpha}-b_{\alpha})\beta_0 + (a_{\alpha}-b_{\alpha})\beta_1 X_{t-1}+ \beta_2 w_{t-1},\\ \nonumber
&=& \gamma_0 + \gamma_1 X_{t-1} + \beta_2 w_{t-1} ,\\ \nonumber
\end{eqnarray}

Lastly, by further relaxing the $\beta_2$ parameter to be a new parameter $\gamma_2$, which allows the time varying $w_t$ component to have its own autoregressive (AR) dynamics with more flexibility, we have:
\begin{equation}\label{wt_addd_x}
w_t =  \gamma_0 + \gamma_1 X_{t-1} + \gamma_2 w_{t-1}.
\end{equation}

Therefore, we propose the Realized-ES-X-CAViaR-X framework as below. The two ``X''s in the model name emphasize the VaR and ES are influenced by the realized measure separately, through two AR specification for $Q_t$ and $w_t$. The last measurement equation ``completes'' the model by regressing the realized measure on the ES, to model their relationship.

\noindent \textbf{Realized-ES-X-CAViaR-X:}
\begin{eqnarray} \label{re_es_caviar_add_x}
&&Q_{t}= \beta_0+ \beta_1 X_{t-1} + \beta_2 Q_{t-1},\\ \nonumber
&& w_t=  \gamma_0 + \gamma_1 X_{t-1} + \gamma_2 w_{t-1} ,\\ \nonumber
&& \text{ES}_t= Q_t - w_t, \\ \nonumber
&&X_t= \xi+\phi |\text{ES}_t|+ \tau_1 \epsilon_t + \tau_2 (\epsilon_t^2-E(\epsilon^2)) + u_t, \,
\end{eqnarray}

where $\gamma_0 \ge 0, \gamma_1 \ge 0, \gamma_2 \ge 0$ to ensure that the VaR and ES estimates do not cross. The definitions of $\epsilon_t$ and $u_t$ are the same as the ones in Realized-ES-CAViaR-Mult and Realized-ES-CAViaR-Add.

In the paper, when needed we use Realized-ES(-X)-CAViaR(-X) as a general name, which includes the proposed Realized-ES-CAViaR-Mult, Realized-ES-CAViaR-Add and Realized-ES-X-CAViaR-X frameworks. The Realized-ES(-X)-CAViaR(-X) frameworks can be extended into other nonlinear models, for example, by choosing the quantile dynamics in Gerlach et al. (2011), which are not explored in the paper.

{\centering
\section{\normalsize LIKELIHOOD AND BAYESIAN ESTIMATION} \label{beyesian_estimation_section}
\par
}
\noindent
\subsection{ES-CAViaR Log Likelihood Function with AL}\label{es_caviar_likelihood_section}


The pseudo-likelihood expression for the quantile and ES equations of the Realized-ES-CAViaR models is the same as that for the ES-CAViaR models in Taylor (2019), being:

\begin{eqnarray}\label{es_caviar_like_equation}
\ell(\mathbf{r};\mathbf{\theta})= \sum_{t=1}^{n} \left( \text{log}  \frac{(\alpha-1)}{\text{ES}_t} + {\frac{(r_t-Q_t)(\alpha-I(r_t\leq Q_t))}{\alpha \text{ES}_t}} \right).
\end{eqnarray}

However, the measurement equation in the Realized-ES-CAViaR model means an added component is needed for a joint pseudo-likelihood function in this framework.

\subsection{Realized-ES(-X)-CAViaR(-X) Log Likelihood}\label{re_es_caviar_likelihood_section}

Because the Realized-ES(-X)-CAViaR(-X) framework has a measurement equation, with $u_t \stackrel{\rm i.i.d.} {\sim} N(0,\sigma_{u}^2)$, the full pseudo-log-likelihood function for Realized-ES(-X)-CAViaR(-X) (as in (\ref{re_es_caviar_ar}) and (\ref{re_es_caviar_exp})) includes two parts and is:
\begin{eqnarray}\label{re_es_caviar_like_equation}
&\ell(\mathbf{r},\mathbf{X};\mathbf{\theta})= \ell(\mathbf{r};\mathbf{\theta})+ \ell(\mathbf{X}|\mathbf{r};\mathbf{\theta})=\\ \nonumber
& \underbrace{\sum_{t=1}^{n} \left( \text{log}  \frac{(\alpha-1)}{\text{ES}_t} + {\frac{(r_t-Q_t)(\alpha-I(r_t\leq Q_t))}{\alpha \text{ES}_t}} \right) }_{\ell (\mathbf{r};\mathbf{\theta})}\\ \nonumber
& \underbrace{-\frac {1}{2} \sum_{t=1}^{n} \big( \text{log} (2 \pi)+ \text{log}(\sigma_{u}^2)+
   u_t^2/\sigma_{u}^2 \big)}_{\ell (\mathbf{X}|\mathbf{r};\mathbf{\theta})} ,\\ \nonumber
\end{eqnarray}
where $u_t= X_t- \xi - \phi |ES_{t}| - \tau_{1} \epsilon_{t} - \tau_{2} (\epsilon_{t}^2-\bar{\epsilon_{t}^2})$, $t=1,\ldots,n$.

In the Realized-GARCH framework, the measurement equation variable $X_t$ contributes to volatility estimation, thus the in-sample and the predictive log-likelihoods are
improved compared to the classical GARCH. We expect that the measurement equation in the Realized-ES(-X)-CAViaR(-X) also facilitates an improved estimate for $ES_{t}$ (and $Q_t$), leading to
more accurate VaR and ES forecasts, compared to the ES-CAViaR models.

\subsection{Maximum Likelihood Estimation}

The pseudo maximum likelihood (ML) approach in Taylor (2019) is adapted and enhanced for the proposed Realized-ES-CAViaR models. Overall, the ML approach includes a four-step process with carefully selected starting values for a robust ``MultiStart'' optimization scheme. In the first step, the quantile equation parameters ($\beta_0,\beta_1,\beta_2$) are estimated separately by optimizing the quantile loss based pseudo-likelihood for a quantile regression.  In the second step, multiple starting values for the measurement equation parameters ($\xi, \phi, \tau_{1}, \tau_{2}, \sigma_{u}$) and ES
component parameter $\gamma_0$ for Realized-ES-CAViaR-Mult, or ($\gamma_0, \gamma_1, \gamma_2)$ for Realized-ES-CAViaR-Add and Realized-ES-X-CAViaR-X, are randomly sampled: 50,000 random candidate starting vectors are
used for Realized-ES-CAViaR-Mult and $10^5$ for Realized-ES-CAViaR-Add and Realized-ES-X-CAViaR-X due to the larger number of parameters involved. In the third step, the estimates for $\beta_0,\beta_1,\beta_2$ in the first step are combined with the randomly sampled candidates in the second step to produce the locally optimum parameter set that maximizes the log-likelihood function (\ref{re_es_caviar_like_equation}). In the fourth step, this locally optimum parameter set is employed as the starting values of the Matlab ``MultiStart'' optimization, which is a robust method of optimization. In our paper, the employed ``MultiStart'' function uses 50 separate starting points, generating 50 local solutions, then the global optimum among these is finally chosen as the ML parameter estimates. As explained in the Matlab documentation: a ``MultiStart'' object contains properties (options) that affect how ``run'' repeatedly runs a local solver to generate a GlobalOptimSolution object. When run, the solver attempts to find multiple local solutions to a problem by starting from various differing points.


\subsection{Bayesian Estimation}

Motivated by Gerlach and Wang (2016), we have also incorporated the Bayesian approach to estimate the proposed models. Given a likelihood function and the specification of a prior distribution, Bayesian approach can be employed to estimate the parameters of the proposed Realized-ES(-X)-CAViaR(-X) models.

An adaptive MCMC method, extended from that in Gerlach and Wang (2016), is employed. Three blocks are employed: $\utwi{\theta_1}=(\beta_0,\beta_1,\beta_2, \phi)$, $\utwi{\theta_2}=(\xi, \tau_{1}, \tau_{2}, \sigma_{u})$, $\utwi{\theta_3}=(\gamma_0, \gamma_1, \gamma_2)$ for Realized-ES-CAViaR-Add and Realized-ES-X-CAViaR-X; and $\utwi{\theta_3}=(\gamma_0)$ for Realized-ES-CAViaR-Mult, via
the motivation that parameters within the same block are more strongly correlated in the posterior (likelihood) than those between blocks.

Priors are chosen to be uninformative over the regions sufficient for positivity and necessary for stationarity,
for example, $\pi(\utwi{\theta})\propto I(A)$, which is a flat prior for $\utwi{\theta}$ over the region $A$. For the Realized-ES-CAViaR-Add and  Realized-ES-X-CAViaR-X models, the region $A$ restricts
$\gamma_0 \ge 0, \gamma_1 \ge 0, \gamma_2 \ge 0$, to ensure positivity of each $w_t$, and also $\gamma_2<1$, which is necessary for stationarity. The restrictions
$\gamma_0 \ge 0, \gamma_1 \ge 0, \gamma_2 \ge 0$ are not strictly necessary for positivity of each $w_t$ term, and in a Bayesian approach these restrictions can be relaxed. However, since in this paper the Bayesian and ML estimation will be directly compared, the restriction is retained here.

In the ``burn-in'' period, an ``epoch'' method as in Chen {et al.} (2017) is employed. For the initial ``epoch'' of the burn-in period, a Metropolis algorithm (Metropolis {et al.}, 1953), employing a mixture of 3 Gaussian proposal distributions with a random walk mean vector, is utilized for each block of parameters. The proposal variance-covariance matrix of each block in each mixture element is $C_i \Sigma$, where $C_1 =1;C_2 =100;C_3 =0.01$, with $\Sigma$ initially set to $\frac{2.38}{\sqrt{(d_i)}}I_{d_i}$, where $d_i$ is the dimension of the parameter block ($i$) and $I_{d_i}$ is the identity matrix of dimension $d_i$. This covariance matrix is subsequently tuned, aiming towards a target acceptance rate of $23.4\%$ (if $d_i>4$, or $35\%$ if $2 \le d_i \le 4$, or $44\%$ if $d_i=1$), as standard, via the algorithm
of Roberts et al. (1997).

In order to enhance the convergence of the chain, at the end of the first epoch, for example, 20,000 iterations, the covariance matrix for each parameter block
is calculated, after discarding (say) the first 2,000 iterations. The covariance matrix is then used in the proposal distribution in the next epoch (of 20,000 iterations).
After each epoch, the standard deviation of each parameter chain in that epoch is calculated and compared to that from the previous epoch. This
process is continued until the mean absolute percentage change of the standard deviation is below a pre-specified threshold, for example, 10\%.
In the empirical study, on average it takes 3--4 epochs to observe this absolute percentage change achieve lower than 10\%; thus, the chains are run for 60,000--80,000 iterations in total
as a burn-in period. A final epoch is run, for say 10,000 iterates, employing an ``independent'' Metropolis Hastings algorithm with a mixture of three Gaussian proposal
distributions for each block. The mean vector for each block is set as the sample mean vector of the last epoch iterates (after discarding the first 2,000 iterates) in the
burn-in period. The proposal variance-covariance matrix in each element is $C_i \Sigma$, where $C_1 =1;C_2 =100;C_3 =0.01$ and $\Sigma$ is the sample covariance matrix of the
last epoch iterates in the burn-in period for that block (after discarding the first 2,000 iterates). Then all the Independent Metropolis Hasting (IMH) iterates (after discarding the first 2,000 iterates)
are employed to calculate the tail risk forecasts, and their posterior mean is used as the final forecast.

The Gelman-Rubin diagnostic (Gelman {et al.}, 2014) is employed to diagnose convergence of the adaptive MCMC method. Further, an effective sample size measure is incorporated to evaluate the efficiency of the MCMC chains (Gelman {et al.}, 2014). To save space, we do not present these efficiency results in detail in the paper. In general,
the Gelman-Rubin statistic is very close to 1 (e.g., $<1.1$) for each parameter, under different block settings, indicating adequate convergence for each parameter using the
adaptive MCMC. By checking more closely the between- and within-chain variances, we observe very small within-chain variances for each parameter, which leads to a close
to 1 Gelman-Rubin statistic and suggests good convergence properties. In addition, the effective sample sizes for all parameters are much larger than the benchmark suggested in Gelman {et al.} (2014), which also highlights the efficiency of the adaptive MCMC algorithm.

In Appendix \ref{simulation_section}, a simulation study is conducted to compare the properties and performance of the Bayesian method and the ML for the Realized-ES(-X)-CAViaR(-X) type models, with respect to parameter estimation
and one-step-ahead VaR and ES forecasting accuracy. The results demonstrate the advantages of employing the MCMC estimator.

{\centering
\section{\normalsize DATA and EMPIRICAL STUDY}\label{data_empirical_section}
\par
}

\subsection{\normalsize REALIZED MEASURES}\label{realized_measure_section}
Various realized measures, including realized variance (RV) and realized range (RR), are incorporated in the proposed Realized-ES(-X)-CAViaR(-X) type models.

To reduce the effect of microstructure noise of realized measures, Martens and van Dijk (2007) present a scaling process inspired by the fact that the daily squared return and range are less affected by microstructure noise than their high frequency counterparts. Therefore, the process can be used to smooth and scale RV and RR, creating less microstructure sensitive measures.

Further, Zhang et al. (2005) propose a sub-sampling process to deal with the micro-structure effects for realized variance (SSRV). The sub-sampling process is applied to RR in Gerlach and Wang (2020b).
The properties of the sub-sampled RR, compared to those of other realized measures, are assessed via simulation under three scenarios in Gerlach and Wang (2020b). Details of employed realized measures are presented in Appendix \ref{realized_measure_section}.

The scaled RV (ScRV), Scaled RR (ScRR), sub-sampled RV (SSRV), and sub-sampled RR (SSRV) are also employed and tested in the proposed frameworks. For example, Realized-ES-CAViaR-Add-RV represents a Realized-ES-CAViaR-Add framework employing RV, and Realized-ES-CAViaR-Mult-RR represents a Realized-ES-CAViaR-Mult framework employing RR. Realized-ES-SSRR-CAViaR-SSRR represents a Realized-ES-X-CAViaR-X framework employing SSRR.

\subsection{Data Description}

Daily and high frequency data, at 1-minute and 5-minute frequency, including open, high, low, and closing prices, are downloaded from
Thomson Reuters Tick History (TRTH) for the period January 2000 to June 2016. Data are collected for seven market indices: S\&P500, NASDAQ (both US), Hang Seng (Hong Kong), FTSE 100 (UK),
DAX (Germany), SMI (Swiss), and ASX200 (Australia); and also for seven individual assets, from the components of the Dow Jones index: CAT (Caterpillar Inc.), KO (Coca-Cola Company Kraft Foods Inc.), GE (General Electric Company), IBM (International Business Machines), JPM (J.P. Morgan Chase \& Company), MMM (3M Company), and XOM (Exxon Mobil Corporation).

\begin{figure}[htp]
     \centering
\includegraphics[width=.8\textwidth]{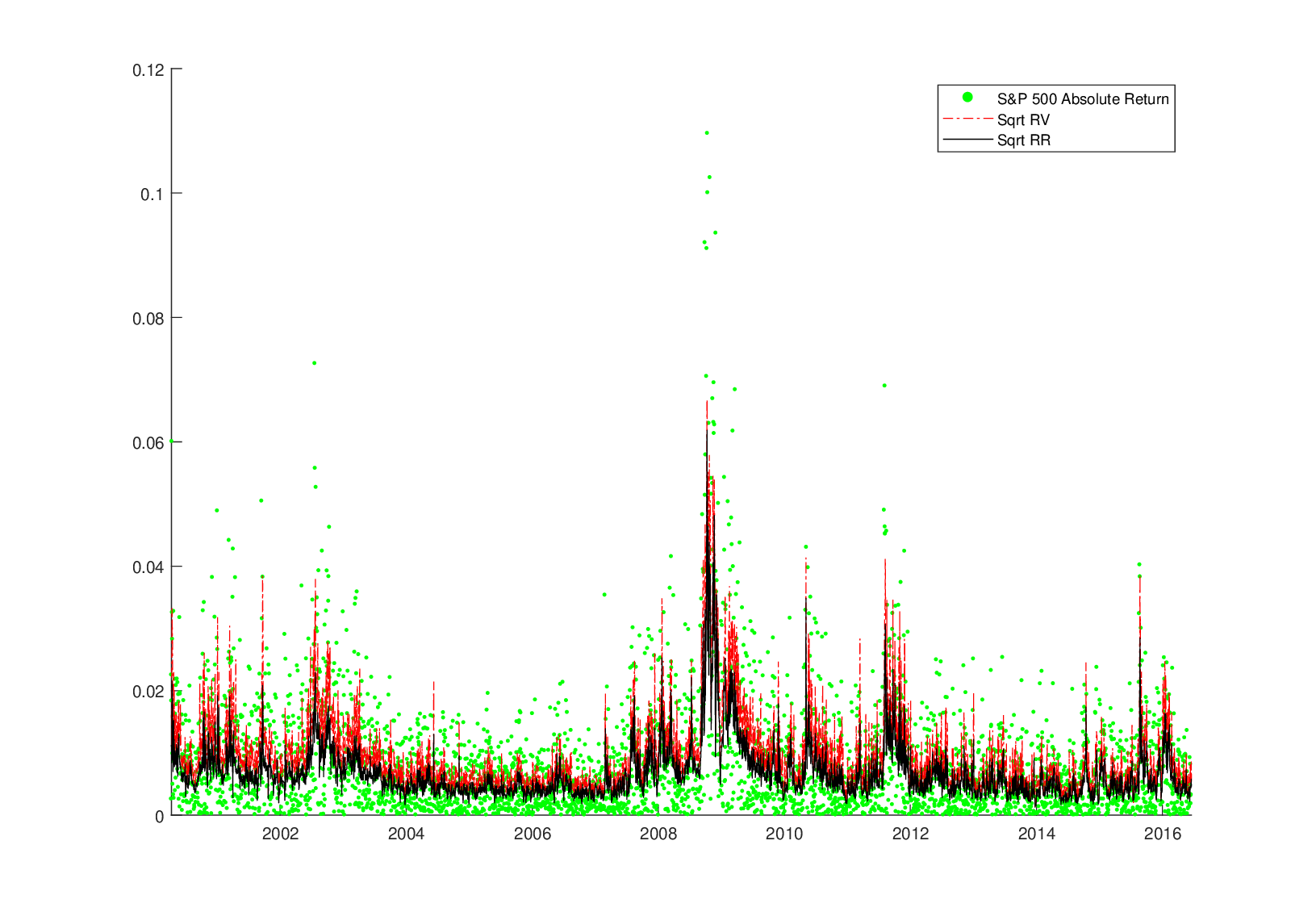}
\caption{\label{Fig2} S\&P 500 absolute return, $\sqrt{RV}$ and $\sqrt{RR}$ plots.}
\end{figure}

The daily price data are used to calculate the daily return, daily range, and daily range plus overnight price jump. Further, the 5-minute prices
are employed to calculate the daily RV, RR, scaled RV, and scaled RR measures, while both 5- and 1-minute data are employed to produce daily sub-sampled RV and RR measures, as in Appendix \ref{realized_measure_section}; $q=66$ is employed for the scaling process, approximately 3 months' daily data. Figure \ref{Fig2} displays the time series plots of the absolute value
of daily returns, $\sqrt{RV}$ and $\sqrt{RR}$ of S\&P 500.  Graphically, the absolute return is a noisy volatility estimator, while both $\sqrt{RV}$ and $\sqrt{RR}$ are less noisy and more efficient estimators. Therefore, incorporating $\sqrt{RV}$ and $\sqrt{RR}$ in the Realized-ES(-X)-CAViaR(-X) type models could potentially improve VaR and ES estimation and forecasting accuracy.

\subsection{Tail Risk Forecasting}

One-step ahead forecasts of both daily VaR and ES are estimated for 14 return series, at quantile levels $\alpha= 1\%, 2.5\%$.

A rolling window, with fixed in-sample data size ($n$), is employed for estimation to produce each of $m$ one-step-ahead forecasts of VaR and ES at $\alpha=2.5\%, 1\%$ in the forecasting period for each series.

The in-sample size $n$ and out-of-sample size $m$ of various datasets can be slightly different, due to different number of trading days for example. The average of the in-sample sizes $n$ and forecast sample sizes $m$ over the indices and assets are given in Table \ref{Summ_var_fore}. In order to assess the model
performance during the period of the global financial crisis (GFC), the forecast period starts from the beginning of 2008. On average, around $\bar{m} = 2110$ one-step-ahead VaR and ES forecasts are generated for each return series from 51 models. These models include the proposed Realized-ES(-X)-CAViaR(-X) models with different input measures of volatility: RV \& RR, scaled RV \& RR, and sub-sampled RV \& RR.

The parametric Realized-GARCH employing the same realized measures are also included for comparison, with both Gaussian (Realized-GARCH-GG) and Student-t (Realized-GARCH-tG) return errors tested.

The ES-CAViaR model of Taylor (2019) with symmetric absolute value and asymmetric slope specifications, and the CARE-SAV model (Taylor, 2008) are also included.

The CARE-AS-X framework (Gerlach et al., 2017), employing the same realized measures, is also evaluated here. In addition, the CARE-AS-X framework incorporating the daily high--low range \& range including the overnight price jump (see Equations (\ref{range_def}) and (\ref{range_o_def}) in Appendix \ref{realized_measure_section} for their definitions) is tested.

In addition, other competing models include: the conventional GARCH (Bollerslev, 1986), EGARCH (Nelson, 1991), and GJR-GARCH (Glosten et al., 1993), all with Student-t errors; a GARCH employing Hansen's skewed-t distribution (Hansen, 1994), three filtered historical simulation (GARCH-HS) type models, employing GARCH-t, EGARCH-t, and GJR-GARCH-t as volatility processes respectively.

The Realized-ES(-X)-CAViaR(-X) and ES-CAViaR models are estimated with adaptive MCMC, and the remaining models are estimated by MLE using the Econometrics toolbox in Matlab (GARCH-t, EGARCH-t, GJR-GARCH-t) or code developed by the authors (GARCH-t-HS, EGARCH-t-HS, GJR-GARCH-t-HS, GARCH-Skew-t, CARE-SAV, CARE-AS-X and Realized-GARCH).

\subsubsection{\normalsize Value-at-Risk}

The VaR violation rate (VRate) is employed to initially assess VaR forecasting accuracy. VRate is simply the proportion of returns that exceed the forecasted VaR in the forecasting period, as in (\ref{varvrate_equation}).
Models with a VRate closest to the nominal quantile levels $\alpha= 1\%, 2.5\%$ are preferred.

\begin{equation}\label{varvrate_equation}
\text{VRate}= \frac{1}{m} \sum_{t=n+1}^{n+m} I(r_t<\text{VaR}_t)\, ,
\end{equation}
where $n$ is the in-sample size and $m$ is the out-of-sample size.

In addition,the standard quantile loss function is also employed to compare
the models for VaR forecast accuracy: the most accurate VaR forecasts should minimize the quantile loss function, given as:
\begin{equation}\label{q_loss}
\frac{1}{m} \sum_{t=n+1}^{n+m}(\alpha-I(r_t<Q_t))(r_t-Q_t) \,\, ,
\end{equation}
where $Q_{n+1},\ldots,Q_{n+m}$ is a series of quantile forecasts at level $\alpha$ for the observations $r_{n+1},\ldots,r_{n+m}$. This quantile loss function is averaged by the out-of-sample size $m$, so that forecasting results across different datasets are comparable.

Table \ref{Summ_var_fore} summarizes the VRates at the 2.5\% and 1\% quantiles over the seven indices and seven assets for each of the 51 models. The ``MAD'' column shows the Mean Absolute Deviation, employing 1\% or 2.5\% as the target VRate, across the seven indices and seven assets respectively. The ``Rank'' column is the average of the ranks, across  seven indices and seven assets, based on the absolute value of the deviation of each VRate from 1\% or 2.5\%. Box indicates the best model, while dashed box indicates the 2nd best model, and blue shading indicates the 3rd best model, according to these measures.

In all the empirical results, the same highlighting rules are applied. Horizontal lines are used to separate models with and without the realized measures. In addition, we place similar type models close to each other, that is, GARCH type, CARE type, ES-CAViaR type, and so on, to make the presentation and comparison clear.

Overall, Table \ref{Summ_var_fore} shows that the proposed Realized-ES(-X)-CAViaR(-X) models produce competitive VRates compared with competing models, and have clearly better performance than the Realized-GARCH type models. By MAD or Rank of (2.5\% or 1\%) VRates across indices or assets, at least one of the proposed Realized-ES(-X)-CAViaR(-X) models ranks as one of the 3 best models, except for the 1\% index study.

However, it can be seen that models without the realized measures, such as GARCH type models, ES-CAViaR-AS models, and CARE-AS models, can produce very good VRate results, for both index and asset on different quantile levels. We now compare these models with our proposed Realized-ES(-X)-CAViaR(-X) models in more detail, with respect to economic efficiency using quantile loss, and provide evidence as to why Realized-ES(-X)-CAViaR(-X) type models are preferred in VaR forecasting.

Table \ref{Summ_quantile_new} shows the average quantile losses for the VaR forecasts for each model, with indices and assets in separate columns. Better models have lower quantile loss. The average rank, based on the ranks of the quantile loss over the 51 models across the seven indices and seven assets, separately, is also included. The results clearly show that the proposed Realized-ES(-X)-CAViaR(-X) type models and Realized-GARCH-tG-SSRV model in general rank better, with lower losses than other models. In particular, the Realized-ES(-X)-CAViaR(-X) models with SSRV and SSRR are preferred. Finally, the quantile losses for models with accurate VRates, that is, GARCH type models, are clearly higher than for the proposed Realized-ES(-X)-CAViaR(-X) models.

\begin{figure}[htp]
     \centering
\includegraphics[width=.8\textwidth]{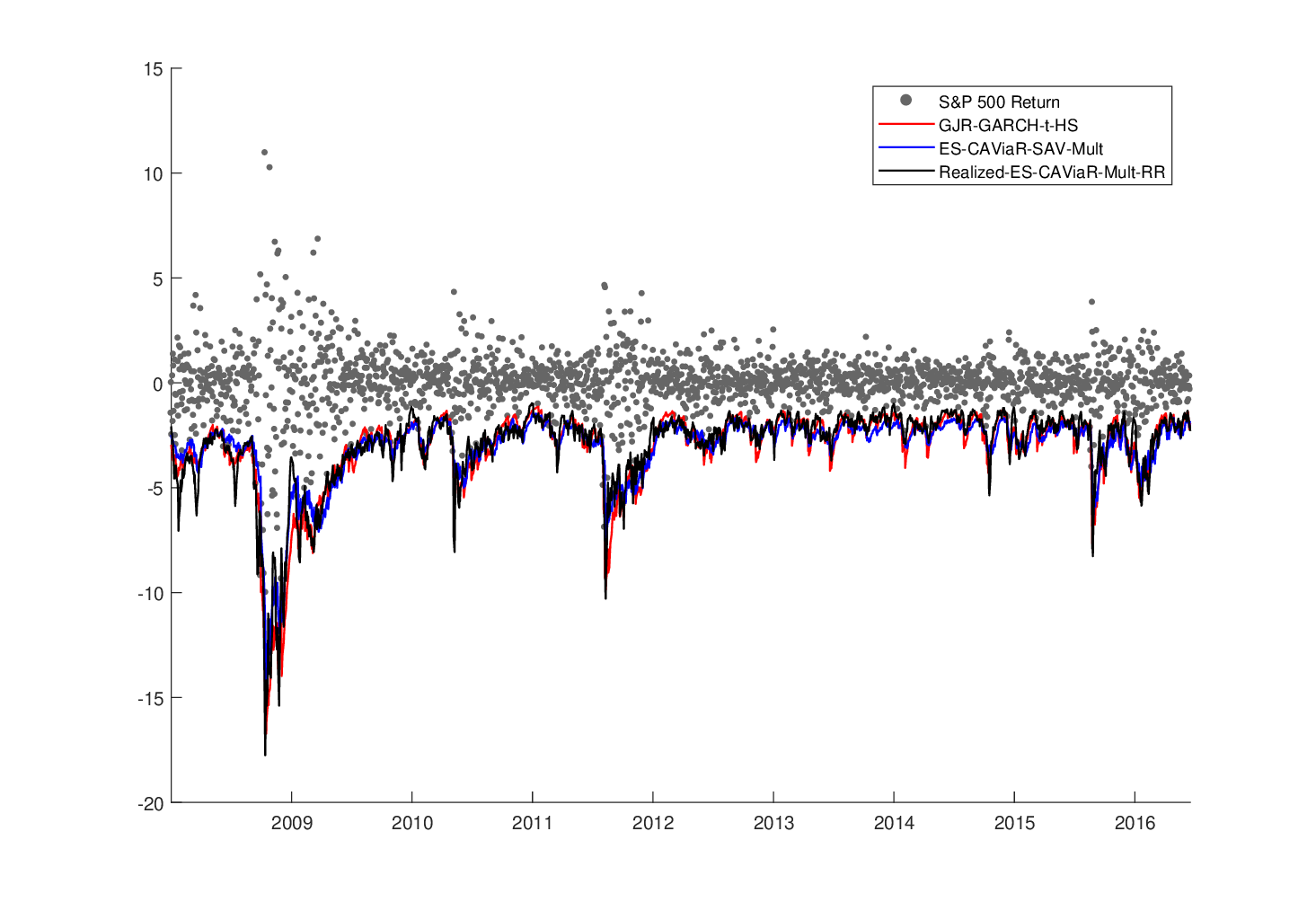}
\caption{\label{var_forecast_fig} S\&P 500 VaR Forecasts with GJR-GARCH-t-HS , ES-CAViaR-SAV-Mult and Realized-ES-CAViaR-Mult-RR.VRates: 1.041\%, 1.278\% and 1.088\%. Quantile losses: 0.0370, 0.0387 and 0.0347.}
\end{figure}
\begin{figure}[htp]
     \centering
\includegraphics[width=.8\textwidth]{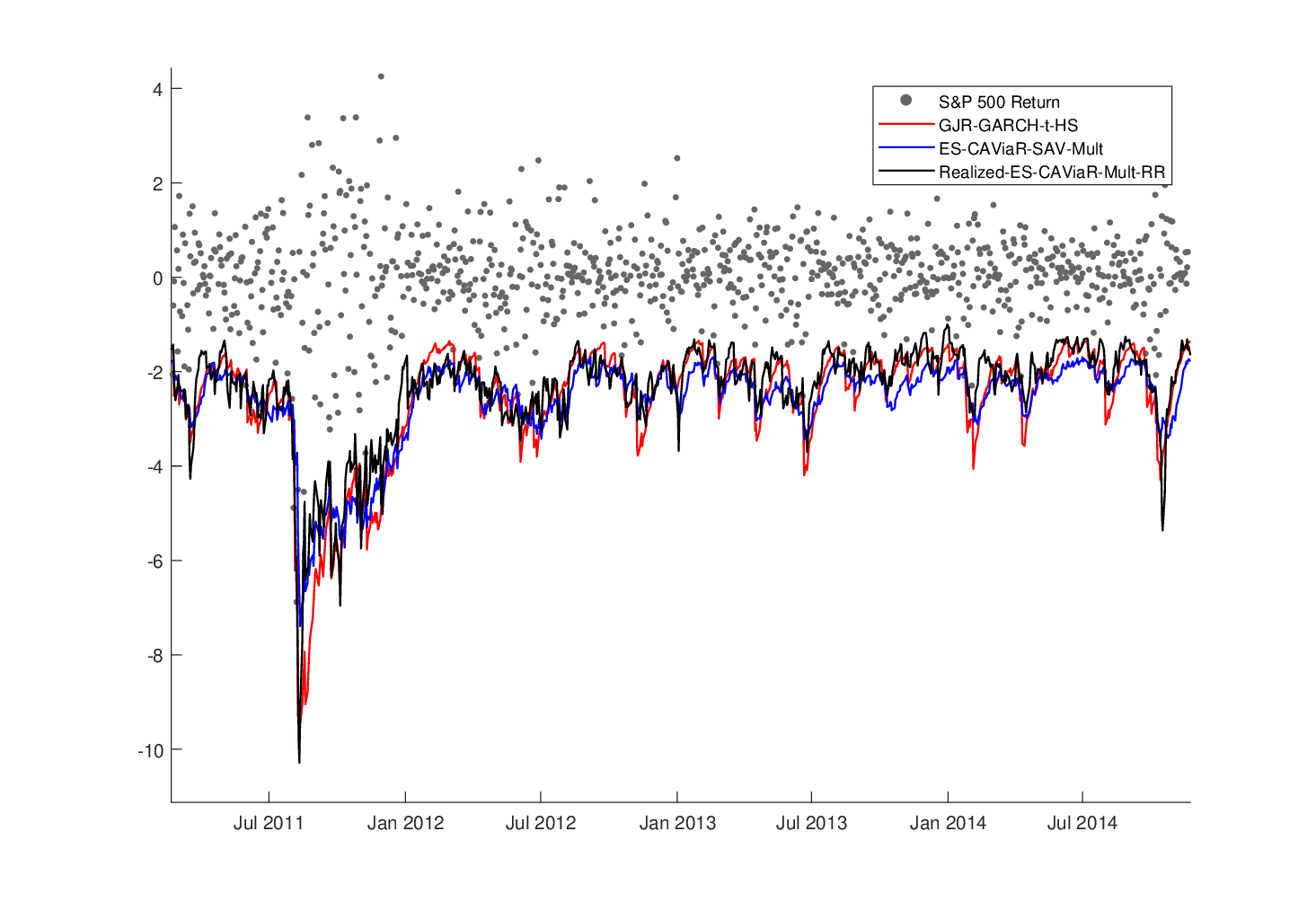}
\caption{\label{var_forecast_fig1} S\&P 500 VaR Forecasts (zoomed in) with GJR-GARCH-t-HS , ES-CAViaR-SAV-Mult and Realized-ES-CAViaR-Mult-RR.
VRates: 1.041\%, 1.278\% and 1.088\%. Quantile losses: 0.0370, 0.0387 and 0.0347.}
\end{figure}

Figures \ref{var_forecast_fig} and \ref{var_forecast_fig1} provide further evidence as to why the proposed Realized-ES-CAViaR type models generate clearly lower quantile
losses, compared to those models also producing VRates close to nominal $1\%$.
Specifically, for the S\&P500 returns, the VRates for the GJR-GARCH-t-HS, ES-CAViaR-SAV-Mult, and Realized-ES-CAViaR-Mult-RR models are 1.041\%, 1.278\%, and 1.088\% respectively: here these 3 models are reasonably close in performance with the GJR-GARCH-t-HS ranked as the best. However, their quantile losses are 0.0370, 0.0387, and 0.0347 respectively, favoring the Realized-ES-CAViaR-Mult-RR model. Through close inspection of Figure \ref{var_forecast_fig1}, it is clear that both ES-CAViaR-SAV-Mult and GJR-GARCH-t-HS generate obviously more extreme (in the negative direction) level of VaR forecasts on most days, compared to the Realized-ES-CAViaR-Mult-RR. This means the capital set aside by financial institutions to cover extreme losses, based on such VaR forecasts, is at a higher level for the GJR-GARCH-t-HS or ES-CAViaR-SAV-Mult models than it is for the Realized-ES-CAViaR-Mult-RR model.

In other words, the Realized-ES-CAViaR-Mult-RR model produces comparatively accurate, with lower loss, VaR forecasts, suggesting that lower amounts of capital are required to protect
against market risk, while simultaneously producing an adequate violation rate, within the Basel Accords framework. For 2113 forecast days for S\&P 500, the forecasts
from Realized-ES-CAViaR-Mult-RR are less extreme than those from ES-CAViaR-SAV-Mult on 1397 days (66.2\%). This suggests a higher level of information (and cost) efficiency regarding risk
levels for the Realized-ES-CAViaR-Mult-RR model, likely coming from the increased statistical efficiency of the realized range series over squared returns, compared to the ES-CAViaR-SAV-Mult
and GJR-GARCH-t-HS models. Since the economic capital is determined by financial institutions' own model and should be
directly proportional to the VaR forecast, the Realized-ES-CAViaR-Mult-RR model is able to decrease the cost capital allocation and increase the profitability of
these institutions, by freeing up part of the regulatory capital from risk coverage into investment, while still providing sufficient and more than
adequate protection against violations. The more accurate and often less extreme VaR forecasts produced by Realized-ES-CAViaR-Mult-RR are particularly
strategically important to decision makers in the financial sector. This extra efficiency is also often observed for proposed Realized-ES-CAViaR-Add and Realized-ES-X-CAViaR-X models over the other markets/assets in this study.

Further, during periods of high volatility, including the GFC, when there is a persistence of extreme returns, the
Realized-ES-CAViaR-Mult-RR VaR forecasts ``recover'' the fastest among the 3 models, as can be seen in Figure \ref{var_forecast_fig1}, in terms of being marginally
the fastest to produce forecasts that again rejoin and follow the tail of the return series. Traditional GARCH models tend to over-react to extreme events and to be
subsequently very slow to recover, due to their oft-estimated very high level of persistence, as discussed in Harvey and Chakravarty (2009). Realized-ES(-X)-CAViaR(-X) models
clearly improve this aspect of performance. Generally, the Realized-ES(-X)-CAViaR(-X) models better describe the dynamics in volatility and tail areas when compared to the traditional
GARCH and the original ES-CAViaR type models, thus largely improving the responsiveness and accuracy of the risk level forecasts, especially following high volatility periods.

Since having a VRate close to the expected level is a necessary but not sufficient condition to guarantee an accurate forecasting model, several quantile accuracy and independence tests are also employed. These tests include: the unconditional coverage (UC) and conditional coverage (CC) tests of Kupiec (1995) and Christoffersen (1998) respectively, as well as the dynamic quantile (DQ using lag equals to 1 and 4 respectively) test of Engle and Manganelli (2004) and the VQR test of Gaglianone et al. (2011).

Table \ref{Summ_bt_new} shows the total number of index or asset series (out of seven, respectively) in which each of the 2.5\% or 1\% VaR forecast models is rejected, conducted at a 5\% significance level. Overall, the Realized-ES-CAViaR models are generally less likely to be rejected by the back tests than other models. Most of the Realized-ES(-X)-CAViaR(-X) type models rank as the top 3 best performing models in each column. In particular, the results of the index clearly show that the Realized-ES(-X)-CAViaR(-X) type models are less likely to be rejected compared with the Realized-GARCH type models.

\subsubsection{\normalsize Expected Shortfall}
The same set of 51 models are employed to generate one-step-ahead forecasts for 1\% and 2.5\% ES over all 14 series in the forecast periods.

To evaluate the ES forecasts, several formal backtests are incorporated, including VaR\&ES joint loss (Fissler and Zeigel, 2016; Taylor, 2017), ES regression backtest (Bayer and Dimitriadis, 2018) and model confidence set (MCS) backtest (Hansen et al., 2011).

\subsubsection{\normalsize VaR\&ES Joint Loss Function}




As discussed in Section \ref{es_caviar_section}, Taylor (2019) shows that the negative logarithm of the likelihood function (\ref{es_caviar_like_equation}) is strictly consistent for $Q_t$ and $ES_t$ considered jointly, and fits into the class of strictly consistent joint functions for VaR and ES developed by Fissler and Zeigel (2016). This loss function as in Equation (\ref{es_caviar_log_score}) is also called the AL log score in Taylor (2019). We use the average joint loss $S = \frac{1}{m} \sum_{t=n+1}^{n+m} S_t$ to formally and jointly assess and compare the VaR and ES forecasts from all models.
\begin{eqnarray}\label{es_caviar_log_score}
S_t(r_t, VaR_t, ES_t) = -\text{log} \left( \frac{\alpha-1}{\text{ES}_t} \right) - {\frac{(r_t-Q_t)(\alpha-I(r_t\leq Q_t))}{\alpha \text{ES}_t}},
\end{eqnarray}

Table \ref{Summ_joint_new} shows the average joint loss $S$ during the forecast period, separately for the indices and assets. The average rank based on the ranks of each model's joint VaR and ES loss $S$ across the indices and assets is also included.

In general, under this measure, the advantages of employing the Realized-ES(-X)-CAViaR(-X) framework are clearly demonstrated. For index, the top 3 ranked models are all from Realized-ES(-X)-CAViaR(-X) family by average loss or average rank. For asset, the Realized-GARCH-tG-SSRV has good performance and appears as the top 3 models for both quantile levels, while the rest of the top performing models are all from Realized-ES(-X)-CAViaR(-X). Again, the Realized-ES(-X)-CAViaR(-X) employing SSRV and SSRV are especially preferred with low joint loss values and good rank.

In addition, Bayer and Dimitriadis (2018) propose a regression based backtest for ES: it regresses the ES forecasts and an intercept term on the returns and tests whether the coefficients are (0,1), for intercept and slope respectively, and is conducted at a 5\% significance level. The results from this backtest, showing the counts of rejections of each model for each test over the 7 indices and 7 assets, are presented in Table \ref{es_back_test}. As can be seen, the test has apparently low power and produces few rejections, except for the Realized-GARCH-GG models, which is the only parametric framework using a Gaussian conditional return distribution. However, in general the proposed Realized-ES(-X)-CaViaR-X class of models is still less (or equally) likely to be rejected by the the test, compared to other models.

\begin{figure}[htp]
     \centering
\includegraphics[width=.8\textwidth]{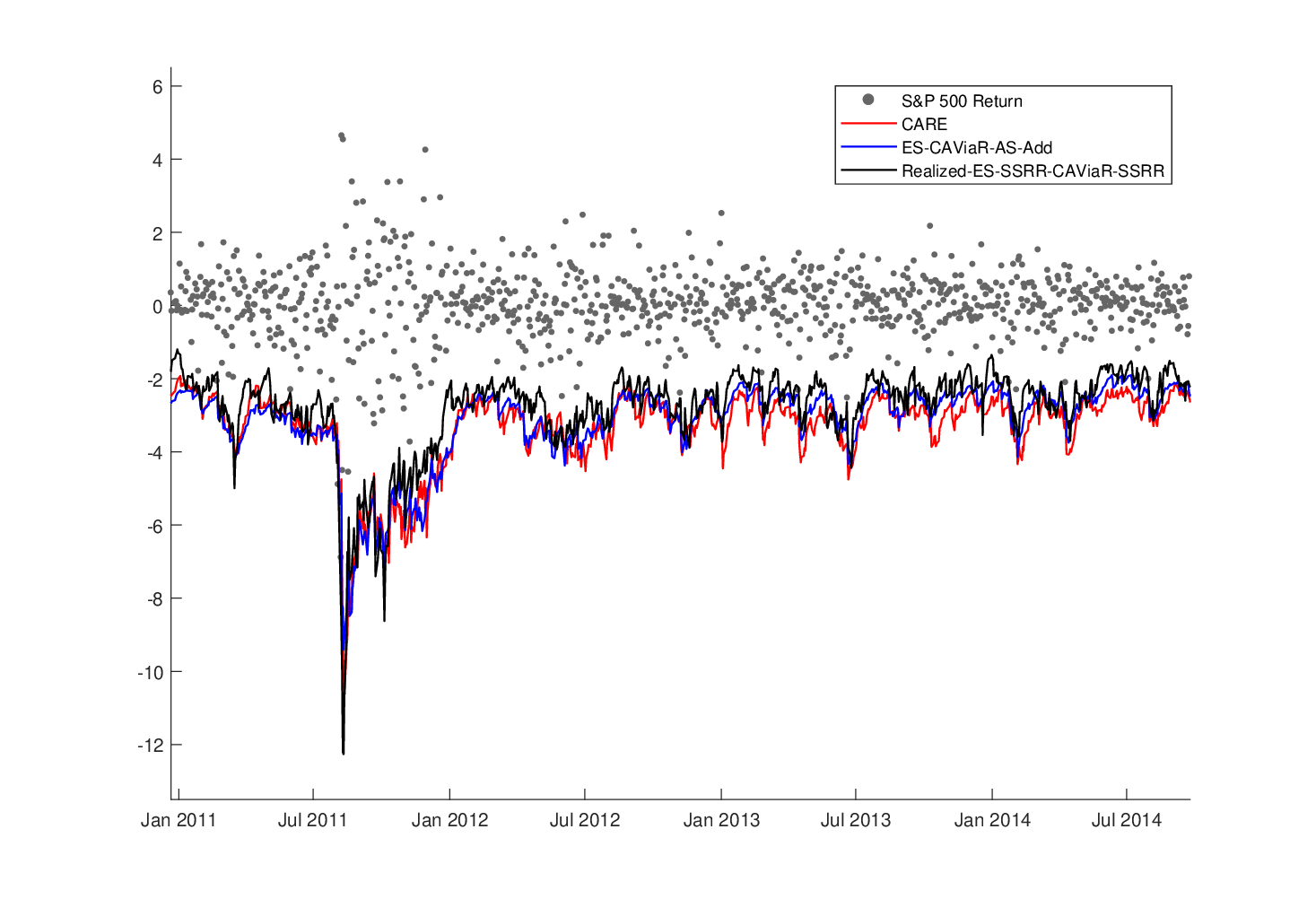}
\caption{\label{Fig_es_fore_zoom_in} S\&P 500 ES Forecasts (zoomed in) with CARE, ES-CAViaR-AS-Add and Realized-ES-SSRR-CAViaR-SSRR. VaR and ES joint loss: 2.2890, 2.2716 and 2.1191.}
\end{figure}

To further demonstrate that extra forecasting efficiency can be gained by employing the proposed Realized-ES(-X)-CAViaR(-X) models, Figure \ref{Fig_es_fore_zoom_in} visualizes the ES forecasts (zoomed in) for S\&P500 from the CARE, ES-CAViaR-AR, and Realized-ES-SSRR-CAViaR-SSRR models. The  VaR and ES joint loss of these three models are 2.2890, 2.2716, and 2.1191 respectively. Figure \ref{Fig_es_fore_zoom_in} shows the cost efficiency gain from Realized-ES(-X)-CAViaR(-X) models is in a similar and even clearer pattern to that from the VaR forecasts. The CARE model produces ES forecasts that are more extreme than the Realized-ES-SSRR-CAViaR-SSRR model's on 1687 days (79.8\%) in the forecast period. In addition, the ES-CAViaR-AS-Add model is more extreme than Realized-ES-SSRR-CAViaR-SSRR on 1363 days (64.5\%).

Therefore, the Realized-ES-X-CAViaR-X employing the sub-sampled realized range apparently improves forecasting efficiency compared with the CARE and ES-CAViaR-AS-Add. Again, this extra efficiency is also frequently observed for the Realized-ES(-X)-CaViaR(-X)
models with other asset return series.

\subsubsection{\normalsize Model Confidence Set}

The MCS, introduced by Hansen et al. (2011), can statistically compare a group of forecast models via a loss function. MCS is applied
to further compare the 27 forecasting models. A MCS is a set of models, constructed such that it contains the
best model with a given level of confidence, selected as 90\% in our paper. The Matlab code for MCS testing is downloaded from ``www.kevinsheppard.com/MFE\_Toolbox'',
adapted to incorporate the VaR and ES joint loss function (\ref{es_caviar_log_score}). Two methods, R and SQ, are employed in the MCS to calculate the test statistics. Specifically, the R method uses the summed absolute values in the calculation, while SQ uses summed squares; more details can be found in Hansen et al. (2011, p. 465).

In Table \ref{Summ_mcs_new}, each column counts the number of series where a model is included in the 90\% MCS, across the seven indices and seven assets and over the 1\% and 2.5\% quantile levels. Box indicates the best model, dashed box indicates the 2nd best model, and blue shading indicates the 3rd best model.

In summary, the proposed Realized-ES(-X)-CAViaR(-X) models are more or equally likely to be included in the MCS compared to other models. The best performing models are Realized-ES(-X)-CAViaR(-X) employing SSRV and SSRR, which are included in the MCS for at least 6 times across both assets and indices and over the 1\% and 2.5\% levels.

Regarding the performance for index, the proposed Realized-ES(-X)-CAViaR(-X) models clearly outperform the other models with realized measures, such as CARE-AS-X and Realized-GARCH.

Regarding the performance for asset, the performance results for the proposed Realized-ES(-X)-CAViaR(-X) models, CARE-AS-X, and Realized-GARCH are close, with most of these models excluded from MCS once across different columns.

Overall, when forecasting 1\% and 2.5\% VaR and ES in 14 financial return series, across several measures and tests for forecast accuracy and model comparison, the performance of the proposed Realized-ES(-X)-CAViaR(-X) models, especially those employing sub-sampled RV and sub-sampled RR, is generally highly favorable.

{\centering
\section{\normalsize CONCLUSION}\label{conclusion_section}
\par
}
\noindent
In this paper, we firstly propose a new semi-parametric realized conditional autoregressive VaR framework called Realized-CAViaR. We show that the proposed Realized-CAViaR includes the Realized-GARCH as a special case, in terms of the VaR estimation and forecast. Secondly, the Realized-CAViaR framework is extended though incorporating various ES components named Realized-ES(-X)-CAViaR(-X). Through incorporating intra-day and high frequency volatility measures, improvements in the out-of-sample forecasting of tail risk measures are observed over a range
of competing models. Specifically, the Realized-ES(-X)-CAViaR(-X) models employing SSRV and SSRR generate the most optimal VaR and ES forecasts in the empirical study
of 14 financial return series. With respect to the backtesting of VaR and ES forecasts, the Realized-ES(-X)-CAViaR(-X) models are also generally less likely to be rejected than their counterparts. Further, the model confidence set results also favor the proposed Realized-ES(-X)-CAViaR(-X) framework.
In addition to being more accurate, the Realized-ES(-X)-CAViaR(-X) models regularly generate less extreme tail risk forecasts, allowing financial institutions to set aside smaller amounts of regulatory capital to protect against market movements.

To conclude, the proposed Realized-ES(-X)-CAViaR(-X) type models, especially those employing sub-sampled RV and sub-sampled RR, should be considered for financial applications in tail
risk forecasting, as they allow financial institutions to more accurately allocate capital under the Basel Capital Accords,
to protect their investments from extreme market movements. This work could be extended by developing asymmetric and non-linear quantile regression specifications, by improving ES component of the model, by using alternative frequencies of observations for the realized measures, and by extending the framework to allow multiple realized measures to appear simultaneously in the model (Hansen and Huang, 2016).

\newpage

\begin{table}[!ht]
\begin{center}
\caption{\label{Summ_var_fore} \small 1\% and 2.5\% VaR forecasting violation rate summary with different models on 7 indices (left panel) and 7 assets (right panel).}\tabcolsep=10pt
\tiny
\begin{tabular}{l p{0.7cm} p{0.7cm}p{0.7cm}p{0.7cm} | p{0.7cm}p{0.7cm}p{0.7cm}p{0.7cm}} \hline
Model                 &1\% MAD              & 1\% Rank     & 2.5\% MAD              & 2.5\% Rank             &1\% MAD              & 1\% Rank     & 2.5\% MAD              & 2.5\% Rank             \\ \hline

GARCH-t&0.632\%&41.4&1.391\%&48.0&0.613\%&30.7&1.329\%&24.0\\
EGARCH-t&0.529\%&37.6&1.161\%&45.3&0.547\%&21.1&1.094\%&18.1\\
GJR-GARCH-t&0.536\%&37.6&1.235\%&45.3&0.566\%&18.4&1.159\%&19.9\\
GARCH-t-HS&0.242\%&16.0&0.289\%&13.3& \fbox{0.135\%}&19.4&0.349\%&18.1\\
EGARCH-t-HS& \cb{0.133\%}&\fbox{6.0}&0.290\%&\cb{13.1}&\cb{0.145\%}&25.9&\dbox{0.274\%}&14.9\\
GJR-GARCH-t-HS&\dbox{0.126\%}&\dbox{6.4}&0.330\%&16.7&0.163\%&17.9&0.330\%&11.6\\
GARCH-Skew-t&0.165\%&14.1&0.329\%&16.3&0.198\%&17.7&0.349\%&\dbox{10.0}\\

CARE&0.272\%&19.4&\dbox{0.269\%}&13.6&0.210\%&23.4&0.321\%&21.0\\
CARE-AS& \fbox{0.125\%}&\cb{8.4}&\fbox{0.268\%}&\dbox{13.0}&0.160\%&\cb{13.3}&0.312\%&13.3\\
CARE-AS-Range&0.262\%&19.6&0.517\%&27.1&0.229\%&31.7&0.470\%&23.4\\
CARE-AS-R-Overnight&0.249\%&18.1&0.336\%&16.6&0.211\%&13.9&0.359\%&22.6\\

ES-CAViaR-SAV-Add&0.282\%&19.9&0.362\%&19.1&0.166\%&18.9&0.347\%&13.6\\
ES-CAViaR-SAV-Mult&0.241\%&16.6&0.308\%&15.9&0.175\%&22.1&0.310\%&17.1\\
ES-CAViaR-AS-Add&0.225\%&15.7&\fbox{0.268\%}&\fbox{11.9}&0.226\%&30.3&0.321\%&14.1\\
ES-CAViaR-AS-Mult&0.158\%&9.4&0.302\%&15.4&0.179\%&22.6&0.359\%&\fbox{9.9}\\

\hline
CARE-AS-RV&0.328\%&21.9&0.482\%&22.3&0.160\%&\dbox{10.6}&0.548\%&29.1\\
CARE-AS-RR&0.193\%&11.4&0.396\%&21.6& \dbox{0.142\%}&19.0&0.413\%&30.6\\
CARE-AS-ScRV&0.275\%&18.3&0.336\%&16.4&0.164\%&19.7&0.415\%&21.3\\
CARE-AS-ScRR&0.302\%&20.6&0.524\%&26.6&0.171\%&17.9&0.536\%&36.9\\
CARE-AS-SSRV&0.262\%&18.0&0.512\%&23.6&0.183\%&15.7&0.451\%&32.6\\
CARE-AS-SSRR&0.233\%&17.6&0.370\%&18.1&0.161\%&18.6&0.347\%&34.1\\

Realized-GARCH-GG-RV&1.048\%&48.1&1.207\%&45.9&0.820\%&45.3&0.943\%&30.1\\
Realized-GARCH-tG-RV&0.437\%&27.7&0.901\%&41.0&0.301\%&29.0&0.716\%&26.0\\
Realized-GARCH-GG-RR&0.878\%&42.0&0.928\%&38.1&0.718\%&47.3&0.774\%&31.3\\
Realized-GARCH-tG-RR&0.439\%&26.6&0.724\%&33.3&0.353\%&26.9&0.538\%&20.9\\
Realized-GARCH-GG-ScRV&0.983\%&48.4&1.072\%&43.6&0.887\%&45.0&0.999\%&26.0\\
Realized-GARCH-tG-ScRV&0.475\%&34.3&0.896\%&38.7&0.398\%&17.6&0.838\%&18.0\\
Realized-GARCH-GG-ScRR&0.929\%&47.3&1.080\%&42.4&0.858\%&47.3&0.942\%&38.1\\
Realized-GARCH-tG-ScRR&0.530\%&37.6&0.877\%&38.4&0.462\%&26.9&0.725\%&30.0\\
Realized-GARCH-GG-SSRV&1.004\%&48.6&1.089\%&42.4&0.877\%&47.6&0.858\%&41.0\\
Realized-GARCH-tG-SSRV&0.638\%&40.6&0.953\%&38.9&0.528\%&18.7&0.734\%&32.9\\
Realized-GARCH-GG-SSRR&0.925\%&45.4&1.029\%&40.6&0.793\%&47.0&0.783\%&33.0\\
Realized-GARCH-tG-SSRR&0.450\%&32.6&0.853\%&37.7&0.426\%&23.0&0.604\%&22.3\\
Realized-ES-CAViaR-Mult-RV&0.414\%&22.9&0.441\%&16.7&0.239\%&15.3&0.329\%&25.0\\
Realized-ES-CAViaR-Mult-RR&0.247\%&19.1&0.328\%&14.7&0.246\%&25.0&0.403\%&24.3\\
Realized-ES-CAViaR-Mult-ScRV&0.281\%&20.4&0.302\%&13.4&0.191\%&16.1&0.311\%&13.9\\
Realized-ES-CAViaR-Mult-ScRR&0.241\%&15.6&0.396\%&21.9&0.220\%&28.3&0.441\%&33.4\\
Realized-ES-CAViaR-Mult-SSRV&0.275\%&20.0&0.438\%&21.9&0.220\%&24.1&0.385\%&30.6\\
Realized-ES-CAViaR-Mult-SSRR&0.240\%&16.7&0.301\%&13.6&0.209\%&25.3&0.347\%&32.3\\
Realized-ES-CAViaR-Add-RV&0.468\%&28.0&0.414\%&13.1&0.239\%&\fbox{8.4}&\fbox{0.262\%}&20.1\\
Realized-ES-CAViaR-Add-RR&0.247\%&16.9&0.299\%&13.4&0.247\%&22.1&0.356\%&24.1\\
Realized-ES-CAViaR-Add-ScRV&0.329\%&24.3&0.295\%&15.0&0.194\%&\fbox{8.4}&\cb{0.302\%}&14.4\\
Realized-ES-CAViaR-Add-ScRR&0.295\%&22.1&0.389\%&20.3&0.248\%&23.9&0.422\%&22.4\\
Realized-ES-CAViaR-Add-SSRV&0.302\%&23.0&0.417\%&21.3&0.239\%&16.3&0.356\%&28.9\\
Realized-ES-CAViaR-Add-SSRR&0.226\%&16.0&\cb{0.281\%}&\dbox{13.0}&0.228\%&22.0&0.318\%&25.6\\
Realized-ES-RV-CAViaR-RV&0.455\%&27.1&0.556\%&23.1&0.229\%&21.3&0.442\%&33.0\\
Realized-ES-RR-CAViaR-RR&0.260\%&18.3&0.356\%&17.4&0.209\%&25.1&0.384\%&28.9\\
Realized-ES-ScRV-CAViaR-ScRV&0.308\%&22.7&0.389\%&20.7&0.210\%&14.4&0.396\%&\cb{11.3}\\
Realized-ES-ScRR-CAViaR-ScRR&0.308\%&22.7&0.491\%&27.1&0.239\%&27.6&0.460\%&36.9\\
Realized-ES-SSRV-CAViaR-SSRV&0.302\%&21.7&0.560\%&28.0&0.230\%&21.0&0.488\%&33.3\\
Realized-ES-SSRR-CAViaR-SSRR&0.247\%&17.4&0.383\%&18.7&0.218\%&26.4&0.366\%&32.6\\ \hline
Out-of-sample size $m$ Avg &  2109.7 & & &  & 2118.7  \\
In-sample size $n$ Avg&  1909.6 & & &   & 1903.0  \\
\hline
\end{tabular}
\end{center}
\emph{Note}:\small  Box indicates the best model, dashed box indicates the 2nd best model, and blue shading indicates the 3rd best model, in each column.
\end{table}

\begin{table}[!ht]
\begin{center}
\caption{\label{Summ_quantile_new} \small  1\% and 2.5\% VaR forecasting quantile loss with different models on 7 indices (left panel) and 7 assets (right panel).}\tabcolsep=10pt
\tiny
\begin{tabular}{l p{0.7cm} p{0.7cm}p{0.7cm}p{0.7cm} | p{0.7cm}p{0.7cm}p{0.7cm}p{0.7cm}}  \hline
Model                 &1\% Avg              & 1\% Rank     & 2.5\% Avg              & 2.5\% Rank             &1\% Avg              & 1\% Rank     & 2.5\% Avg              & 2.5\% Rank             \\ \hline

GARCH-t&0.0410&44.1&0.0863&47.6&0.0560&41.4&0.1121&43.7\\
EGARCH-t&0.0394&31.3&0.0837&33.4&0.0551&36.1&0.1103&35.4\\
GJR-GARCH-t&0.0396&34.4&0.0837&33.7&0.0549&35.4&0.1103&34.9\\
GARCH-t-HS&0.0407&43.0&0.0852&42.3&0.0561&43.0&0.1125&45.6\\
EGARCH-t-HS&0.0393&26.3&0.0829&27.0&0.0549&35.6&0.1107&36.4\\
GJR-GARCH-t-HS&0.0395&33.0&0.0830&29.9&0.0548&36.3&0.1107&37.1\\
GARCH-Skew-t&0.0406&41.7&0.0852&42.9&0.0560&41.1&0.1121&44.0\\

CARE&0.0417&44.4&0.0860&46.4&0.0584&45.1&0.1146&45.7\\
CARE-AS&0.0400&35.0&0.0839&34.9&0.0574&42.3&0.1118&39.0\\
CARE-AS-Range&0.0390&23.0&0.0828&26.6&0.0555&36.0&0.1093&29.6\\
CARE-AS-R-Overnight&0.0388&21.3&0.0824&22.9&0.0565&39.6&0.1102&33.1\\

ES-CAViaR-SAV-Add&0.0410&42.1&0.0856&43.6&0.0570&44.7&0.1127&45.7\\
ES-CAViaR-SAV-Mult&0.0409&42.0&0.0855&44.3&0.0566&41.0&0.1129&44.4\\
ES-CAViaR-AS-Add&0.0393&27.9&0.0829&28.3&0.0557&40.1&0.1104&36.1\\
ES-CAViaR-AS-Mult&0.0392&26.6&0.0829&27.4&0.0558&39.7&0.1107&38.0\\

\hline
CARE-AS-RV&0.0401&35.9&0.0829&21.4&0.0540&34.9&0.1087&27.9\\
CARE-AS-RR&0.0386&18.3&0.0823&18.6&0.0546&35.3&0.1078&26.7\\
CARE-AS-ScRV&0.0396&30.6&0.0829&26.1&0.0552&36.7&0.1094&32.6\\
CARE-AS-ScRR&0.0388&22.6&0.0820&20.7&0.0549&34.0&0.1087&27.3\\
CARE-AS-SSRV&0.0386&21.0&\cb{0.0813}&13.0&0.0537&29.1&0.1079&24.4\\
CARE-AS-SSRR&0.0393&24.1&0.0816&14.3&0.0537&32.4&0.1080&26.7\\

Realized-GARCH-GG-RV&0.0413&35.3&0.0847&34.0&0.0530&29.0&0.1071&24.6\\
Realized-GARCH-tG-RV&0.0398&24.1&0.0841&29.6&0.0518&14.9&0.1072&26.1\\
Realized-GARCH-GG-RR&0.0401&26.7&0.0834&25.0&0.0526&26.7&0.1067&22.1\\
Realized-GARCH-tG-RR&0.0388&16.6&0.0830&22.3&0.0518&21.3&0.1065&21.3\\
Realized-GARCH-GG-ScRV&0.0403&37.4&0.0839&35.7&0.0529&24.3&0.1072&21.6\\
Realized-GARCH-tG-ScRV&0.0392&26.4&0.0835&33.1&0.0524&20.1&0.1073&22.1\\
Realized-GARCH-GG-ScRR&0.0397&33.9&0.0827&27.9&0.0530&27.9&0.1070&22.1\\
Realized-GARCH-tG-ScRR&0.0386&23.0&0.0823&25.0&0.0519&16.9&0.1068&21.7\\
Realized-GARCH-GG-SSRV&0.0393&27.3&0.0822&20.7&0.0513&13.4&\dbox{0.1047}&\dbox{7.6}\\
Realized-GARCH-tG-SSRV&\fbox{0.0380}&\fbox{10.3}&0.0817&17.1&\fbox{0.0505}&\fbox{5.0}&\fbox{0.1044}&\fbox{5.7}\\
Realized-GARCH-GG-SSRR&0.0394&22.1&0.0827&20.3&0.0525&25.7&0.1067&21.7\\
Realized-GARCH-tG-SSRR&\cb{0.0383}&12.6&0.0824&18.7&0.0518&20.0&0.1065&20.1\\
Realized-ES-CAViaR-Mult-RV&0.0403&27.4&0.0843&29.3&0.0517&15.4&0.1069&23.1\\
Realized-ES-CAViaR-Mult-RR&0.0389&19.0&0.0824&16.4&0.0515&14.6&0.1062&18.1\\
Realized-ES-CAViaR-Mult-ScRV&0.0395&28.3&0.0834&33.0&0.0522&21.3&0.1078&25.0\\
Realized-ES-CAViaR-Mult-ScRR&0.0388&21.4&0.0820&20.7&0.0521&17.6&0.1072&18.4\\
Realized-ES-CAViaR-Mult-SSRV&\dbox{0.0382}&\dbox{12.1}&\dbox{0.0812}&\dbox{10.9}&\cb{0.0509}&\cb{9.0}&0.1056&12.6\\
Realized-ES-CAViaR-Mult-SSRR&0.0384&\cb{12.3}&0.0816&\cb{12.0}&0.0512&13.3&0.1061&18.1\\
Realized-ES-CAViaR-Add-RV&0.0398&25.5&0.0835&27.6&0.0513&9.3&0.1063&18.3\\
Realized-ES-CAViaR-Add-RR&0.0388&19.2&0.0824&19.0&0.0515&14.3&0.1059&14.7\\
Realized-ES-CAViaR-Add-ScRV&0.0394&27.4&0.0836&34.7&0.0520&16.4&0.1072&20.3\\
Realized-ES-CAViaR-Add-ScRR&0.0388&21.1&0.0818&17.7&0.0518&14.1&0.1067&15.4\\
Realized-ES-CAViaR-Add-SSRV&\dbox{0.0382}&\dbox{12.1}&\fbox{0.0809}&\fbox{8.6}&\dbox{0.0506}&\dbox{7.4}&\cb{0.1051}&\cb{8.7}\\
Realized-ES-CAViaR-Add-SSRR&\cb{0.0383}&12.6&0.0817&15.0&0.0512&12.3&0.1058&14.9\\
Realized-ES-RV-CAViaR-RV&0.0398&23.2&0.0833&24.7&0.0520&18.3&0.1071&24.4\\
Realized-ES-RR-CAViaR-RR&0.0389&20.6&0.0824&17.6&0.0520&20.4&0.1066&23.0\\
Realized-ES-ScRV-CAViaR-ScRV&0.0395&27.9&0.0835&33.3&0.0526&23.9&0.1079&26.0\\
Realized-ES-ScRR-CAViaR-ScRR&0.0389&23.6&0.0822&20.0&0.0525&22.3&0.1075&21.6\\
Realized-ES-SSRV-CAViaR-SSRV&0.0384&14.4&0.0816&15.0&0.0514&12.0&0.1058&13.0\\
Realized-ES-SSRR-CAViaR-SSRR&0.0385&13.5&0.0819&16.0&0.0519&19.0&0.1063&18.9\\
\hline
\end{tabular}
\end{center}
\emph{Note}:\small Box indicates the best model, dashed box indicates the 2nd best model, and blue shading indicates the 3rd best model, in each column.
\end{table}

\begin{table}[hbt!]
\begin{center}
\caption{\label{Summ_bt_new} \small Total number of 1\% and 2.5 \% VaR forecasting rejections with UC, CC, DQ1, and DQ4 and VQR tests for different models on 7 indices and 7 assets. All tests conducted at 5\% significance level.}\tabcolsep=10pt
\tiny
\begin{tabular}{lcc|ccccccccccc} \hline
& \multicolumn{2}{c}{Index} & \multicolumn{2}{c}{Asset} & \\
Model& 1\% Total& 2.5\% Total & 1\% Total & 2.5\% Total\\
\hline

GARCH-t&7&7&5&6\\
EGARCH-t&6&6&5&\dbox{2}\\
GJR-GARCH-t&6&6&3&4\\
GARCH-t-HS&\dbox{2}&\cb{2}&4&4\\
EGARCH-t-HS&\dbox{2}&\cb{2}&4&\dbox{2}\\
GJR-GARCH-t-HS&\fbox{1}&\cb{2}&\dbox{1}&\cb{3}\\
GARCH-Skew-t&3&\cb{2}&3&4\\

CARE&4&\cb{2}&4&\cb{3}\\
CARE-AS&\fbox{1}&\cb{2}&3&\dbox{2}\\
CARE-AS-Range&5&3&\cb{2}&\cb{3}\\
CARE-AS-R-Overnight&\cb{3}&\dbox{1}&\dbox{1}&\dbox{2}\\

ES-CAViaR-SAV-Add&4&4&3&4\\
ES-CAViaR-SAV-Mult&\cb{3}&3&\cb{2}&\cb{3}\\
ES-CAViaR-AS-Add&\fbox{1}&\cb{2}&3&\dbox{2}\\
ES-CAViaR-AS-Mult&\dbox{2}&\cb{2}&4&\fbox{1}\\

\hline
CARE-AS-RV&\cb{3}&\cb{2}&3&\cb{3}\\
CARE-AS-RR&4&\fbox{0}&\cb{2}&\cb{3}\\
CARE-AS-ScRV&\dbox{2}&3&4&\cb{3}\\
CARE-AS-ScRR&\cb{3}&3&3&4\\
CARE-AS-SSRV&4&3&4&5\\
CARE-AS-SSRR&4&\fbox{0}&\cb{2}&4\\
Realized-GARCH-GG-RV&7&6&6&\cb{3}\\
Realized-GARCH-tG-RV&\cb{3}&6&\cb{2}&\cb{3}\\
Realized-GARCH-GG-RR&6&7&7&5\\
Realized-GARCH-tG-RR&5&5&\dbox{1}&4\\
Realized-GARCH-GG-ScRV&7&6&5&\dbox{2}\\
Realized-GARCH-tG-ScRV&5&6&\cb{2}&\dbox{2}\\
Realized-GARCH-GG-ScRR&7&6&6&\cb{3}\\
Realized-GARCH-tG-ScRR&7&7&4&\dbox{2}\\
Realized-GARCH-GG-SSRV&7&6&6&5\\
Realized-GARCH-tG-SSRV&7&6&\dbox{1}&4\\
Realized-GARCH-GG-SSRR&7&6&7&5\\
Realized-GARCH-tG-SSRR&5&7&3&4\\
Realized-ES-CAViaR-Mult-RV&\dbox{2}&4&\cb{2}&4\\
Realized-ES-CAViaR-Mult-RR&\dbox{2}&\cb{2}&\cb{2}&\dbox{2}\\
Realized-ES-CAViaR-Mult-ScRV&\dbox{2}&\cb{2}&\dbox{1}&\dbox{2}\\
Realized-ES-CAViaR-Mult-ScRR&\cb{3}&\cb{2}&5&\dbox{2}\\
Realized-ES-CAViaR-Mult-SSRV&4&\cb{2}&4&5\\
Realized-ES-CAViaR-Mult-SSRR&\cb{3}&\cb{2}&4&4\\
Realized-ES-CAViaR-Add-RV&\dbox{2}&3&\cb{2}&\dbox{2}\\
Realized-ES-CAViaR-Add-RR&\dbox{2}&\dbox{1}&3&\cb{3}\\
Realized-ES-CAViaR-Add-ScRV&\cb{3}&\dbox{1}&\fbox{0}&\fbox{1}\\
Realized-ES-CAViaR-Add-ScRR&\cb{3}&\dbox{1}&\cb{2}&\fbox{1}\\
Realized-ES-CAViaR-Add-SSRV&5&\cb{2}&3&\cb{3}\\
Realized-ES-CAViaR-Add-SSRR&\dbox{2}&\dbox{1}&3&\cb{3}\\
Realized-ES-RV-CAViaR-RV&\dbox{2}&4&\dbox{1}&4\\
Realized-ES-RR-CAViaR-RR&\cb{3}&\cb{2}&3&4\\
Realized-ES-ScRV-CAViaR-ScRV&4&3&\fbox{0}&\dbox{2}\\
Realized-ES-ScRR-CAViaR-ScRR&\cb{3}&4&3&\cb{3}\\
Realized-ES-SSRV-CAViaR-SSRV&5&3&\dbox{1}&4\\
Realized-ES-SSRR-CAViaR-SSRR&4&\cb{2}&\dbox{1}&5\\

\hline
\end{tabular}
\end{center}
\emph{Note}:\small Box indicates the best model, dashed box indicates the 2nd best model, and blue shading indicates the 3rd best model, in each column. The number is the lower the better.

\end{table}

\begin{table}[!ht]
\begin{center}
\caption{\label{Summ_joint_new} \small 2.5\% and 1\% VaR \& ES forecasting joint loss with different models on 7 indices (left panel) and 7 assets (right panel).}\tabcolsep=10pt
\tiny
\begin{tabular}{l p{0.7cm} p{0.7cm}p{0.7cm}p{0.7cm} | p{0.7cm}p{0.7cm}p{0.7cm}p{0.7cm}}  \hline
Model                 &1\% Avg              & 1\% Rank     & 2.5\% Avg              & 2.5\% Rank             &1\% Avg              & 1\% Rank     & 2.5\% Avg              & 2.5\% Rank             \\ \hline

GARCH-t&2.3491&46.1&2.1960&48.9&2.6490&39.9&2.4293&45.0\\
EGARCH-t&2.3282&40.6&2.1632&39.9&2.6375&34.7&2.4131&34.9\\
GJR-GARCH-t&2.3177&37.4&2.1616&41.0&2.6354&34.7&2.4148&37.6\\
GARCH-t-HS&2.3287&41.3&2.1680&42.3&2.6497&40.0&2.4318&45.4\\
EGARCH-t-HS&2.3104&35.3&2.1445&32.9&2.6371&35.6&2.4160&36.9\\
GJR-GARCH-t-HS&2.3005&32.4&2.1416&33.1&2.6351&35.4&2.4178&38.9\\
GARCH-Skew-t&2.3253&40.3&2.1698&44.0&2.6464&38.3&2.4284&43.3\\

CARE&2.3667&44.7&2.1852&46.7&2.6783&42.0&2.4370&42.4\\
CARE-AS&2.3222&37.0&2.1548&35.6&2.6813&41.0&2.4208&36.1\\
CARE-AS-Range&2.2772&21.6&2.1316&24.1&2.6398&36.0&2.3984&32.9\\
CARE-AS-R-Overnight&2.2728&21.4&2.1306&23.1&2.6686&39.3&2.4046&32.7\\

ES-CAViaR-SAV-Add&2.3466&44.7&2.1757&45.4&2.6594&40.3&2.4303&45.4\\
ES-CAViaR-SAV-Mult&2.3421&43.1&2.1744&44.3&2.6511&37.9&2.4264&43.1\\
ES-CAViaR-AS-Add&2.3038&33.0&2.1410&29.9&2.6497&40.3&2.4108&37.0\\
ES-CAViaR-AS-Mult&2.2971&30.6&2.1399&29.9&2.6425&38.0&2.4101&36.6\\

\hline
CARE-AS-RV&2.3039&31.4&2.1345&21.0&2.6055&32.4&2.3839&26.3\\
CARE-AS-RR&2.2714&20.7&2.1224&17.7&2.6578&34.1&2.3750&23.6\\
CARE-AS-ScRV&2.2893&27.0&2.1317&24.1&2.6347&36.4&2.3932&31.3\\
CARE-AS-ScRR&2.2706&20.0&2.1189&17.6&2.6489&34.1&2.3817&25.3\\
CARE-AS-SSRV&2.2806&24.1&2.1107&\cb{11.3}&2.6027&24.7&2.3748&22.6\\
CARE-AS-SSRR&2.3004&27.7&2.1126&12.3&2.6031&31.0&2.3795&27.0\\
Realized-GARCH-GG-RV&2.3485&34.7&2.1630&36.1&2.6323&38.9&2.3870&32.0\\
Realized-GARCH-tG-RV&2.2882&21.7&2.1424&27.1&2.5682&15.9&2.3739&24.4\\
Realized-GARCH-GG-RR&2.3115&27.7&2.1393&26.0&2.6340&36.3&2.3832&30.1\\
Realized-GARCH-tG-RR&2.2626&14.7&2.1252&20.1&2.5675&20.0&2.3665&18.3\\
Realized-GARCH-GG-ScRV&2.3320&40.3&2.1572&38.7&2.6199&35.0&2.3868&30.3\\
Realized-GARCH-tG-ScRV&2.2798&22.3&2.1420&31.9&2.5744&19.3&2.3775&22.9\\
Realized-GARCH-GG-ScRR&2.3113&34.4&2.1372&29.4&2.6332&37.0&2.3848&29.0\\
Realized-GARCH-tG-ScRR&2.2644&19.0&2.1234&22.3&2.5674&16.3&2.3679&18.4\\
Realized-GARCH-GG-SSRV&2.3109&33.3&2.1350&28.3&2.6111&32.1&2.3696&18.6\\
Realized-GARCH-tG-SSRV&2.2572&12.3&2.1199&19.4&\dbox{2.5492}&\dbox{7.3}&\fbox{2.3523}&\fbox{5.6}\\
Realized-GARCH-GG-SSRR&2.3029&28.6&2.1349&26.3&2.6366&37.3&2.3855&29.7\\
Realized-GARCH-tG-SSRR&2.2567&13.7&2.1239&22.1&2.5691&18.4&2.3684&18.7\\
Realized-ES-CAViaR-Mult-RV&2.3042&25.3&2.1473&26.1&2.5610&13.3&2.3655&16.1\\
Realized-ES-CAViaR-Mult-RR&2.2651&15.7&2.1186&15.0&2.5632&14.3&2.3627&15.7\\
Realized-ES-CAViaR-Mult-ScRV&2.2835&23.9&2.1369&28.0&2.5662&15.7&2.3756&24.3\\
Realized-ES-CAViaR-Mult-ScRR&2.2642&16.7&2.1156&16.4&2.5684&16.4&2.3694&17.9\\
Realized-ES-CAViaR-Mult-SSRV&\fbox{2.2538}&11.7&\dbox{2.1064}&\dbox{9.0}&\cb{2.5505}&\fbox{7.1}&\dbox{2.3552}&\cb{9.4}\\
Realized-ES-CAViaR-Mult-SSRR&\dbox{2.2555}&\dbox{10.6}&\cb{2.1096}&\cb{11.3}&2.5616&13.1&2.3632&14.7\\
Realized-ES-CAViaR-Add-RV&2.2926&24.1&2.1388&27.3&2.5562&9.1&2.3651&15.3\\
Realized-ES-CAViaR-Add-RR&2.2666&17.6&2.1228&18.1&2.5623&13.1&2.3624&15.0\\
Realized-ES-CAViaR-Add-ScRV&2.2863&26.7&2.1430&32.9&2.5630&14.1&2.3750&23.3\\
Realized-ES-CAViaR-Add-ScRR&2.2678&20.3&2.1165&17.9&2.5661&14.7&2.3678&15.4\\
Realized-ES-CAViaR-Add-SSRV&\cb{2.2556}&13.3&\fbox{2.1062}&\fbox{7.9}&\fbox{2.5490}&\cb{8.1}&2.3562&\dbox{8.6}\\
Realized-ES-CAViaR-Add-SSRR&2.2573&12.0&2.1127&13.3&2.5615&12.7&2.3628&14.7\\
Realized-ES-RV-CAViaR-RV&2.2888&22.4&2.1329&22.6&2.5617&14.0&2.3670&18.4\\
Realized-ES-RR-CAViaR-RR&2.2647&16.7&2.1197&15.9&2.5660&16.1&2.3652&18.3\\
Realized-ES-ScRV-CAViaR-ScRV&2.2838&23.7&2.1368&28.7&2.5709&19.6&2.3768&25.9\\
Realized-ES-ScRR-CAViaR-ScRR&2.2664&19.3&2.1177&17.1&2.5716&19.4&2.3704&21.3\\
Realized-ES-SSRV-CAViaR-SSRV&2.2558&\cb{11.6}&2.1100&12.0&2.5525&8.7&\cb{2.3561}&11.4\\
Realized-ES-SSRR-CAViaR-SSRR&2.2558&\fbox{10.3}&2.1126&13.7&2.5653&16.3&2.3646&18.0\\
\hline
\end{tabular}
\end{center}
\emph{Note}:\small Box indicates the best model, dashed box indicates the 2nd best model, and blue shading indicates the 3rd best model, in each column.
\end{table}

\begin{table}[hbt!]
\begin{center}
\caption{\label{es_back_test} \small Total number of 1\% and 2.5 \% ES regression backtest rejections for different models on 7 indices and 7 assets. All tests conducted at 5\% significance level.}\tabcolsep=10pt
\tiny
\begin{tabular}{lcc|ccccccccccc} \hline
& \multicolumn{2}{c}{Index} & \multicolumn{2}{c}{Asset} & \\
Model& 1\% Index Total& 2.5\% Index Total & 1\% Asset Total & 2.5\% Asset Total\\
\hline
GARCH-t&0&4&1&0\\
EGARCH-t&0&1&1&1\\
GJR-GARCH-t&0&2&1&1\\
GARCH-t-HS&0&0&2&0\\
EGARCH-t-HS&0&0&2&1\\
GJR-GARCH-t-HS&0&1&2&1\\
GARCH-Skew-t&0&0&1&0\\

CARE&0&0&0&0\\
CARE-AS&0&0&0&0\\
CARE-AS-Range&0&0&1&0\\
CARE-AS-R-Overnight&0&0&0&0\\

ES-CAViaR-SAV-Add&0&0&0&0\\
ES-CAViaR-SAV-Mult&0&0&0&0\\
ES-CAViaR-AS-Add&0&0&0&0\\
ES-CAViaR-AS-Mult&0&0&1&0\\

\hline
CARE-AS-RV&1&1&0&0\\
CARE-AS-RR&0&0&1&0\\
CARE-AS-ScRV&0&0&0&0\\
CARE-AS-ScRR&0&0&0&0\\
CARE-AS-SSRV&0&0&0&0\\
CARE-AS-SSRR&0&0&1&0\\
Realized-GARCH-GG-RV&5&6&6&6\\
Realized-GARCH-tG-RV&0&1&0&0\\
Realized-GARCH-GG-RR&5&6&6&6\\
Realized-GARCH-tG-RR&0&2&2&0\\
Realized-GARCH-GG-ScRV&6&7&4&4\\
Realized-GARCH-tG-ScRV&0&0&1&0\\
Realized-GARCH-GG-ScRR&6&6&6&6\\
Realized-GARCH-tG-ScRR&0&1&1&0\\
Realized-GARCH-GG-SSRV&6&6&6&6\\
Realized-GARCH-tG-SSRV&0&1&2&0\\
Realized-GARCH-GG-SSRR&6&6&6&6\\
Realized-GARCH-tG-SSRR&0&1&2&0\\
Realized-ES-CAViaR-Mult-RV&1&1&0&0\\
Realized-ES-CAViaR-Mult-RR&0&0&1&0\\
Realized-ES-CAViaR-Mult-ScRV&0&0&0&0\\
Realized-ES-CAViaR-Mult-ScRR&0&0&0&0\\
Realized-ES-CAViaR-Mult-SSRV&0&0&0&0\\
Realized-ES-CAViaR-Mult-SSRR&0&0&0&0\\
Realized-ES-CAViaR-Add-RV&1&1&0&0\\
Realized-ES-CAViaR-Add-RR&0&1&0&0\\
Realized-ES-CAViaR-Add-ScRV&0&0&0&0\\
Realized-ES-CAViaR-Add-ScRR&0&0&1&1\\
Realized-ES-CAViaR-Add-SSRV&0&0&0&0\\
Realized-ES-CAViaR-Add-SSRR&0&1&0&0\\
Realized-ES-RV-CAViaR-RV&1&1&0&0\\
Realized-ES-RR-CAViaR-RR&1&0&0&0\\
Realized-ES-ScRV-CAViaR-ScRV&0&0&0&0\\
Realized-ES-ScRR-CAViaR-ScRR&0&0&0&0\\
Realized-ES-SSRV-CAViaR-SSRV&0&0&0&0\\
Realized-ES-SSRR-CAViaR-SSRR&0&0&0&0\\
\hline
\end{tabular}
\end{center}
\emph{Note}:\small The number is the lower the better.
\end{table}

\begin{table}[!ht]
\begin{center}
\caption{\label{Summ_mcs_new} \small 90\% model confidence set with R and SQ methods for different models on 7 indices and 7 assets.}\tabcolsep=10pt
\tiny
\begin{tabular}{lcccc | ccccccc} \hline
& \multicolumn{4}{c}{Index} & \multicolumn{4}{c}{Asset} & \\
Model                 &1\% R              & 1\% SQ     & 2.5\% R              & 2.5\% SQ             &1\% R              & 1\% SQ     & 2.5\% R              & 2.5\% SQ             \\ \hline
GARCH-t&3&4&2&1&4&\cb{5}&\cb{5}&4\\
EGARCH-t&\dbox{6}&\cb{5}&\cb{5}&\cb{5}&\dbox{6}&\dbox{6}&\dbox{6}&4\\
GJR-GARCH-t&\dbox{6}&\cb{5}&3&4&4&\cb{5}&4&4\\
GARCH-t-HS&\cb{5}&\cb{5}&2&4&3&\cb{5}&4&4\\
EGARCH-t-HS&\dbox{6}&\cb{5}&\cb{5}&\cb{5}&\dbox{6}&\dbox{6}&\cb{5}&\cb{5}\\
GJR-GARCH-t-HS&\dbox{6}&\cb{5}&\dbox{6}&\dbox{6}&3&\cb{5}&4&4\\
GARCH-Skew-t&\cb{5}&\cb{5}&3&4&4&\dbox{6}&4&4\\

CARE&3&\cb{5}&2&3&\cb{5}&4&\cb{5}&4\\
CARE-AS&\cb{5}&\cb{5}&\cb{5}&\dbox{6}&4&4&\cb{5}&4\\
CARE-AS-Range&\fbox{7}&\dbox{6}&\dbox{6}&\dbox{6}&\dbox{6}&\dbox{6}&\dbox{6}&\cb{5}\\
CARE-AS-R-Overnight&\dbox{6}&\fbox{7}&\cb{5}&\cb{5}&\cb{5}&4&\cb{5}&4\\

ES-CAViaR-SAV-Add&2&\cb{5}&2&3&4&\cb{5}&\cb{5}&4\\
ES-CAViaR-SAV-Mult&4&\cb{5}&2&3&\cb{5}&4&\cb{5}&\cb{5}\\
ES-CAViaR-AS-Add&\dbox{6}&\cb{5}&4&\cb{5}&4&4&\cb{5}&\cb{5}\\
ES-CAViaR-AS-Mult&\dbox{6}&\cb{5}&\dbox{6}&\dbox{6}&4&\cb{5}&\cb{5}&\cb{5}\\

\hline
CARE-AS-RV&4&4&4&4&\cb{5}&\fbox{7}&\fbox{7}&\dbox{6}\\
CARE-AS-RR&\dbox{6}&\dbox{6}&\cb{5}&\dbox{6}&\cb{5}&\fbox{7}&\fbox{7}&\fbox{7}\\
CARE-AS-ScRV&\cb{5}&\cb{5}&\cb{5}&\cb{5}&\cb{5}&\dbox{6}&\fbox{7}&\dbox{6}\\
CARE-AS-ScRR&\dbox{6}&\dbox{6}&\dbox{6}&\cb{5}&\dbox{6}&\fbox{7}&\fbox{7}&\fbox{7}\\
CARE-AS-SSRV&\dbox{6}&\cb{5}&\dbox{6}&\dbox{6}&\dbox{6}&\fbox{7}&\fbox{7}&\dbox{6}\\
CARE-AS-SSRR&4&\fbox{7}&\cb{5}&\dbox{6}&\dbox{6}&\fbox{7}&\fbox{7}&\dbox{6}\\

Realized-GARCH-GG-RV&\cb{5}&4&4&4&\dbox{6}&\fbox{7}&\fbox{7}&\dbox{6}\\
Realized-GARCH-tG-RV&\cb{5}&4&4&4&\dbox{6}&\fbox{7}&\fbox{7}&\dbox{6}\\
Realized-GARCH-GG-RR&\cb{5}&4&\cb{5}&\cb{5}&\dbox{6}&\fbox{7}&\dbox{6}&\dbox{6}\\
Realized-GARCH-tG-RR&\dbox{6}&\cb{5}&\dbox{6}&\dbox{6}&\dbox{6}&\fbox{7}&\dbox{6}&\dbox{6}\\
Realized-GARCH-GG-ScRV&\cb{5}&4&4&4&\cb{5}&\fbox{7}&\fbox{7}&\dbox{6}\\
Realized-GARCH-tG-ScRV&\cb{5}&\cb{5}&4&\cb{5}&\dbox{6}&\fbox{7}&\fbox{7}&\dbox{6}\\
Realized-GARCH-GG-ScRR&\cb{5}&\cb{5}&\cb{5}&\dbox{6}&\dbox{6}&\fbox{7}&\dbox{6}&\dbox{6}\\
Realized-GARCH-tG-ScRR&\fbox{7}&\dbox{6}&\dbox{6}&\dbox{6}&\fbox{7}&\fbox{7}&\fbox{7}&\dbox{6}\\
Realized-GARCH-GG-SSRV&\cb{5}&\cb{5}&\cb{5}&\dbox{6}&\dbox{6}&\fbox{7}&\fbox{7}&\dbox{6}\\
Realized-GARCH-tG-SSRV&\fbox{7}&\dbox{6}&\dbox{6}&\dbox{6}&\fbox{7}&\fbox{7}&\fbox{7}&\fbox{7}\\
Realized-GARCH-GG-SSRR&\dbox{6}&\cb{5}&\cb{5}&\cb{5}&\dbox{6}&\dbox{6}&\dbox{6}&\dbox{6}\\
Realized-GARCH-tG-SSRR&\dbox{6}&\dbox{6}&\dbox{6}&\dbox{6}&\dbox{6}&\fbox{7}&\dbox{6}&\dbox{6}\\
Realized-ES-CAViaR-Mult-RV&4&\cb{5}&3&4&\fbox{7}&\fbox{7}&\fbox{7}&\fbox{7}\\
Realized-ES-CAViaR-Mult-RR&\dbox{6}&\dbox{6}&\cb{5}&\cb{5}&\dbox{6}&\fbox{7}&\fbox{7}&\fbox{7}\\
Realized-ES-CAViaR-Mult-ScRV&\cb{5}&4&3&4&\fbox{7}&\fbox{7}&\fbox{7}&\dbox{6}\\
Realized-ES-CAViaR-Mult-ScRR&\dbox{6}&\dbox{6}&\cb{5}&\dbox{6}&\dbox{6}&\fbox{7}&\fbox{7}&\fbox{7}\\
Realized-ES-CAViaR-Mult-SSRV&\fbox{7}&\dbox{6}&\fbox{7}&\fbox{7}&\fbox{7}&\fbox{7}&\fbox{7}&\fbox{7}\\
Realized-ES-CAViaR-Mult-SSRR&\dbox{6}&\fbox{7}&\dbox{6}&\dbox{6}&\fbox{7}&\fbox{7}&\fbox{7}&\fbox{7}\\
Realized-ES-CAViaR-Add-RV&\dbox{6}&\cb{5}&3&3&\fbox{7}&\fbox{7}&\fbox{7}&\fbox{7}\\
Realized-ES-CAViaR-Add-RR&\dbox{6}&\cb{5}&\cb{5}&\cb{5}&\dbox{6}&\fbox{7}&\fbox{7}&\fbox{7}\\
Realized-ES-CAViaR-Add-ScRV&\cb{5}&4&3&3&\fbox{7}&\fbox{7}&\fbox{7}&\dbox{6}\\
Realized-ES-CAViaR-Add-ScRR&\dbox{6}&\dbox{6}&\cb{5}&\dbox{6}&\dbox{6}&\fbox{7}&\fbox{7}&\dbox{6}\\
Realized-ES-CAViaR-Add-SSRV&\fbox{7}&\dbox{6}&\fbox{7}&\fbox{7}&\fbox{7}&\fbox{7}&\fbox{7}&\fbox{7}\\
Realized-ES-CAViaR-Add-SSRR&\dbox{6}&\dbox{6}&\dbox{6}&\dbox{6}&\fbox{7}&\fbox{7}&\fbox{7}&\dbox{6}\\
Realized-ES-RV-CAViaR-RV&\dbox{6}&\cb{5}&\cb{5}&4&\fbox{7}&\fbox{7}&\fbox{7}&\fbox{7}\\
Realized-ES-RR-CAViaR-RR&\dbox{6}&\cb{5}&\cb{5}&\cb{5}&\dbox{6}&\fbox{7}&\dbox{6}&\fbox{7}\\
Realized-ES-ScRV-CAViaR-ScRV&\cb{5}&4&3&4&\fbox{7}&\fbox{7}&\fbox{7}&\dbox{6}\\
Realized-ES-ScRR-CAViaR-ScRR&\cb{5}&\cb{5}&\cb{5}&\dbox{6}&\dbox{6}&\fbox{7}&\fbox{7}&\fbox{7}\\
Realized-ES-SSRV-CAViaR-SSRV&\fbox{7}&\dbox{6}&\dbox{6}&\dbox{6}&\fbox{7}&\fbox{7}&\fbox{7}&\fbox{7}\\
Realized-ES-SSRR-CAViaR-SSRR&\dbox{6}&\fbox{7}&\dbox{6}&\dbox{6}&\fbox{7}&\fbox{7}&\fbox{7}&\fbox{7}\\

\hline
\end{tabular}
\end{center}
\emph{Note}:\small Box indicates the best model, dashed box indicates the 2nd best model, and blue shading indicates the 3rd best model, in each column. The number is the higher the better.
\end{table}

\clearpage
\section*{References}
\addcontentsline{toc}{section}{References}
\begin{description}

\item Acerbi, C., Tasche, D. 2002.  Expected shortfall: A natural coherent alternative to value at risk. \emph{Economic Notes}, 31(2), 379–388.

\item Andersen, T. G., Bollerslev, T. 1998. Answering the skeptics: Yes, standard volatility models do provide accurate forecasts. \emph{International Economic Review}, 885-905.

\item Andersen, T. G., Bollerslev, T., Diebold, F. X., Labys, P. 2003. Modeling and forecasting realized volatility.
    \emph{Econometrica}, 71(2), 579-625.

\item Artzner, P., Delbaen, F., Eber, J.M., Heath, D. 1997. Thinking coherently.  \emph{Risk}, 10, 68-71.

\item Artzener, P., Delbaen, F., Eber, J.M., Heath, D. 1999. Coherent measures of risk.  \emph{Mathematical Finance}, 9, 203-228.

\item Bayer, S., Dimitriadis, T. 2018. Regression based expected shortfall backtesting. arXiv preprint arXiv:1801.04112.

\item Basel Committee on Banking Supervision 2019. Minimum Capital Requirements for Market Risk.  Bank for International Settlements.

\item Bollerslev, T. 1986. Generalized autoregressive conditional heteroskedasticity. \emph{Journal of Econometrics}, 31, 307-327.

\item Chen, W., Peters, G., Gerlach, R., Sisson, S. 2017. Dynamic quantile function models. arXiv:1707.02587.

\item Christensen, K., Podolskij, M. 2007. Realized range-based estimation of integrated variance. \emph{Journal of Econometrics},
141(2), 323-349.

\item Christoffersen, P. 1998. Evaluating interval forecasts. \emph{International Economic Review}, 39, 841-862.

\item Contino, C., Gerlach, R. 2017. Bayesian tail‐risk forecasting using realized GARCH. \emph{Applied Stochastic Models in Business and Industry}, 33(2), 213-236.

\item Creal, D., Koopman, S.J., Lucas, A. 2013. Generalized autoregressive score models with applications. \emph{Journal of Applied Econometrics}, 28(5), 777-795.

\item Engle, R. F. 1982. Autoregressive conditional heteroskedasticity with estimates of the variance of United Kingdom
inflations. \emph{Econometrica}, 50, 987-1007.

\item Engle, R. F., Manganelli, S. 2004. CAViaR: Conditional autoregressive value at risk
by regression quantiles. \emph{Journal of Business and Economic Statistics}, 22, 367-381.

\item Feller, W. 1951. The asymptotic distribution of the range of sums of random variables. \emph{Annals of Mathematical Statistics}, 22, 427-32.

\item Fissler, T., Ziegel, J. F. 2016. Higher order elicibility and Osband's principle. \emph{The Annals of Statistics}, 44(4), 1680-1707.

\item Gaglianone, W. P., Lima, L. R., Linton, O., Smith, D. R. 2011. Evaluating value-
at-risk models via quantile regression. \emph{Journal of Business and Economic Statistics},
29, 150-160.

\item Garman, M. B., Klass, M. J. 1980. On the estimation of security price volatilities from historical data.
\emph{The Journal of Business}, 67-78.

\item Gelman, A., Carlin, J.B., Stern, H.S., Rubin, D.B. 2014. \emph{Bayesian Data Analysis (Vol. 2)}. CRC Press, Boca Raton, FL.

\item Gerlach, R., Chen, C.W.S. 2016. Bayesian expected shortfall forecasting incorporating the intraday range.
\emph{Journal of Financial Econometrics}, 14(1), 128-158.

\item  Gerlach, R., Chen, C.W., Chan, N.Y. 2011. Bayesian time-varying quantile forecasting for value-at-risk in financial markets. \emph{Journal of Business \& Economic Statistics}, 29(4), 481-492.

\item Gerlach, R., Wang, C. 2016.  Forecasting risk via realized GARCH, incorporating the realized range.
\emph{Quantitative Finance}, 16(4), 501-511.

\item Gerlach, R. and Wang, C., 2020a. Semi-parametric dynamic asymmetric Laplace models for tail risk forecasting, incorporating realized measures. \emph{International Journal of Forecasting}, 36(2), 489-506.

\item Gerlach, R. and Wang, C., 2020b. Bayesian Semi-parametric Realized Conditional Autoregressive Expectile Models for Tail Risk Forecasting. \emph{Journal of Financial Econometrics}, in press.

\item Glosten, L.R., Jagannathan, R., Runkle, D.E. 1993. On the relation between the expected value and the volatility of the nominal excess return on stocks. \emph{The Journal of Finance}, 48(5), 1779-1801.

\item Hansen, B. E. 1994. Autoregressive conditional density estimation. \emph{International Economic Review}, 35, 705-730.


\item Hansen, P. R., Huang, Z., Shek, H. H. 2012. Realized GARCH: A joint model for returns and realized measures of volatility. \emph{Journal of Applied Econometrics}, 27(6), 877-906.

\item Hansen, P. R., Huang, Z. 2016. Exponential GARCH modeling with realized measures of volatility. \emph{Journal of Business \& Economic Statistics}, 34(2), 269-287.

\item Hansen, P.R., Lunde, A., Nason, J.M. 2011. The model confidence set. \emph{Econometrica}, 79(2), 453-497.

\item Harvey, A.C., Chakravarty, T. 2009. Beta-t-EGARCH. Working paper. Earlier version appeared in 2008 as a Cambridge Working paper in Economics, CWPE 0840.


\item Harvey, A.C. 2013. Dynamic Models for Volatility and Heavy Tails: With Applications to Financial and Economic Time Series (Vol. 52). Cambridge University Press, Cambridge.

\item Koenker, R., Machado, J.A. 1999. Goodness of fit and related inference processes for quantile regression. \emph{Journal of the American Statistical Association}, 94(448), 1296-1310.

\item Kupiec, P. H. 1995. Techniques for verifying the accuracy of risk measurement models. \emph{The Journal of Derivatives}, 3, 73-84.

\item Martens, M., van Dijk, D. 2007. Measuring volatility with the realized range. \emph{Journal of Econometrics}, 138(1), 181-207.

\item Metropolis, N., Rosenbluth, A. W., Rosenbluth, M. N., Teller, A. H., Teller, E. 1953. Equation of state calculations by fast
computing machines. \emph{J. Chem. Phys}, 21, 1087-1092.

\item Nelson, D. B. 1991. Conditional heteroskedasticity in asset returns: A new approach. \emph{Econometrica}, 59, 347-370.

\item Parkinson, M. 1980. The extreme value method for estimating the variance of the rate of return. \emph{Journal of Business},
53(1), 61.

\item Patton, A.J., Ziegel, J.F., Chen, R. 2019. Dynamic semiparametric models for expected shortfall (and value-at-risk). \emph{Journal of Econometrics}, 211(2), pp.388-413.

\item Roberts, G. O., Gelman, A., Gilks, W. R. 1997. Weak convergence and optimal scaling of random walk Metropolis algorithms.
    \emph{The Annals of Applied Probability}, 7(1), 110-120.

\item Taylor, J. 2008. Estimating value at risk and expected shortfall using expectiles. \emph{Journal of Financial Econometrics}, 6, 231-252.

\item Taylor, J. 2019. Forecasting value at risk and expected shortfall using a semiparametric approach based on the asymmetric Laplace distribution. \emph{Journal of Business and Economic Statistics}, 37(1), 121-133.

\item Watanabe, T. 2012. Quantile forecasts of financial returns using Realized GARCH Models. \emph{Japanese Economic Review}, 63(1),
    68-80.

\item Zhang, L., Mykland, P. A., A\"{i}t-Sahalia, Y. 2005. A tale of two time scales.  \emph{Journal of the American Statistical
    Association}, 100(472).

\end{description}

\clearpage
\appendixtitleon
\appendixtitletocon
\begin{appendices}

{\centering
\section{\normalsize SIMULATION STUDY}\label{simulation_section}
\par
}

\noindent
A simulation study on $\alpha = 1\%$  quantile level is conducted to compare the properties and performance of the Bayesian method and the Maximum Likelihood for the Realized-ES-CAViaR type models, with respect to parameter estimation
and one-step-ahead VaR and ES forecasting accuracy. Both the mean and Root Mean Square Error (RMSE) values are calculated over the replicated datasets for the MCMC and ML methods, to illustrate their
respective bias and precision. $\alpha = 1\%$ is chosen, which is a more extreme case than $\alpha = 2.5\%$ to test the performance of MCMC and ML more strictly.

As discussed in Section \ref{realized_cav_model_section}, in the Abs-Realized-GARCH framework (\ref{rgarch}) with a parametric conditional error distribution, for example, normal distribution, the ratio between ES and VaR is fixed and able to be known exactly,
that is, $Q_{t}= \sigma_t \Phi^{-1}(\alpha)$, and $\text{ES}_{t}= - \sigma_t \frac{\phi(\Phi^{-1}(\alpha))}{\alpha} $, where $\Phi^{}$ is the standard Normal cdf, $\phi()$ is standard Normal pdf, and
$\sigma_t$ is the volatility. Thus, $\frac{\text{ES}_{t}}{Q_{t}} = \frac{-\phi(\Phi^{-1}(\alpha))}{\alpha \Phi^{-1}(\alpha)} = 1.1457$ when $\alpha= 1\%$. Therefore, the Realized-ES-CAViaR-Mult model is analogous to the dynamics of VaR and ES in a parametric Abs-Realized-GARCH framework. We can thus know the analogous true parameters of the Realized-ES-CAViaR-Mult model exactly, when simulating from a parametric Abs-Realized-GARCH model.

More specifically, suppose 1000 simulated datasets with sample size $n=1900$ are generated from the following Abs-Realized-GARCH model, as in (\ref{r_garch_simu}). The sample size $n=1900$ is chosen to be close to the in-sample size in the empirical study, that is, $n$ in Table \ref{Summ_var_fore}. The values of parameters in (\ref{r_garch_simu}) are also chosen based on the estimated parameters values in the empirical study.

\begin{eqnarray} \label{r_garch_simu}
&&r_t= \sigma_t z_t, \\ \nonumber
&&\sigma_t= 0.02 + 0.10 X_{t-1}+ 0.85 \sigma_t,  \\ \nonumber
&&X_t= 0.1+0.9 \sigma_t-0.02 z_t  + 0.02 (z_t^{2}-1) +u_t, \\  \nonumber
&& z_t \stackrel{\rm i.i.d.} {\sim} N(0,1), u_t \stackrel{\rm i.i.d.} {\sim} N(0,0.3^2).\\  \nonumber
\end{eqnarray}

In order to calculate the corresponding Realized-ES-CAViaR true parameter values, a mapping from the square root Realized-GARCH to the
Realized-ES-CAViaR can be derived. With $Q_{t}= \sigma_t \Phi^{-1}(\alpha)$ and  $\text{ES}_{t}= - \sigma_t \frac{\phi(\Phi^{-1}(\alpha))}{\alpha} $, we have
$\sigma_t =\frac{Q_{t}} {\Phi^{-1}(\alpha)}=  -\frac{\alpha \text{ES}_{t}}{\phi(\Phi^{-1}(\alpha))}$. Further, with $z_t \stackrel{\rm i.i.d.} {\sim} N(0,1)$, we have
$\epsilon_{t}= \frac{r_t} {Q_{t}} = \frac{r_t} {\sigma_t \Phi^{-1}(\alpha)} = \frac{z_t} {\Phi^{-1}(\alpha)}$.

Substituting back into GARCH and measurement equations of Abs-Realized-GARCH fraemework (\ref{r_garch_simu}), the corresponding Realized-ES-CAViaR (without ES component) specification can be written as:
\begin{eqnarray}
&& Q_{t}= 0.02 \Phi^{-1}(\alpha) + 0.10 \Phi^{-1}(\alpha) X_{t-1}+ 0.85 Q_{t-1} ,\\ \nonumber
&& X_t= 0.1 + \frac{0.9 \alpha}{\phi(\Phi^{-1}(\alpha))} | \text{ES}_{t} |-0.02\Phi^{-1}(\alpha)\epsilon_{t} + 0.02 \Phi^{-1}(\alpha)^2 (\epsilon_{t}^2-  \frac{1} {\Phi^{-1}(\alpha)^2} ) + u_t,  \nonumber
\end{eqnarray}
allowing true parameter values to be calculated or read off. These true values are presented in Tables \ref{simu_table_mult}, \ref{simu_table_add}, and \ref{simu_table_add_x}.

In addition, the true one-step-ahead $\alpha$ level VaR and ES forecasts can be calculated exactly via (\ref{r_garch_simu}); that is, $Q_{t+1}= \sigma_{t+1} \Phi^{-1}(\alpha)$,
and $\text{ES}_{t+1}= -\sigma_{t+1}  \frac{\phi(\Phi^{-1}(\alpha))}{\alpha} $. For each simulated dataset, the true values of $Q_{n+1}$ and $\text{ES}_{n+1}$ are recorded;
the averages of these, over the 1000 datasets, are given in the ``True'' column of Tables \ref{simu_table_mult}, \ref{simu_table_add} and \ref{simu_table_add_x}, respectively.

Regarding the parameters in the ES component, for the Realized-ES-CAViaR-Mult model the implied value of $\gamma_0$ can be solved for each $t$, through solving $\frac{\text{ES}_{t}}{Q_{t}} = \frac{-\phi(\Phi^{-1}(\alpha))}{\alpha \Phi^{-1}(\alpha)} = 1.1457=(1+exp(\gamma_0))$. The ``True'' value of $\gamma_0$, which equals $-1.9264$, is presented in the ``True'' column of Table \ref{simu_table_mult}. However, in the additive version of the proposed models, that is, Realized-ES-CAViaR-Add and Realized-ES-X-CAViaR-X, the true values of $\gamma_0, \gamma_1, \gamma_2$ cannot be solved exactly, nor in a similar manner to that for the Realized-ES-CAViaR-Mult model. This is because the additive version of the proposed models, that is, Realized-ES-CAViaR-Add and Realized-ES-X-CAViaR-X, allows the observed input series, for example, lagged returns or realized measures, to separately influence VaR and ES. This breaks the fixed relationship between VaR and ES that pertains in a parametric Abs-Realized-GARCH type model. For example, when deriving the $w_t$ component Equation (\ref{wt_addd_x}) of the Realized-ES-X-CAViaR-X, we have relaxed the parameter $\beta_2$ to $\gamma_2$, to allow $w_t$ to have its own autoregressive structure with realized measures $X_t$ as input. For the Realized-ES-CAViaR-Add in Model \ref{re_es_caviar_ar}, the $w_t$ uses the presented autoregressive structure (with return as input) only when return violates the quantile, that is, $r_t \le Q_t$. It is an open question as to what type of parametric model allows separate and unlinked dynamics on VaR and ES, that is, the ones in Realized-ES-CAViaR-Add and Realized-ES-X-CAViaR-X that could potentially help estimate and forecast VaR and ES more accurately.

Therefore, in the simulation study (using Abs-Realized-GARCH model (\ref{r_garch_simu}) with Normal error as data generating process) we firstly focus on Realized-ES-CAViaR-Mult models, the true parameter values of which can all be calculated exactly, as presented above. In the Realized-ES-CAViaR-Add and Realized-ES-X-CAViaR-X models, as discussed, the true values of $\gamma_0, \gamma_1, \gamma_2$ cannot be solved exactly. However, we still test the performance of MCMC and ML on estimating other parameters and producing the one-step-ahead VaR and ES forecast. The Realized-ES-CAViaR-Mult,  Realized-ES-CAViaR-Add, and Realized-ES-X-CAViaR-X models are fit to each dataset, once using MCMC and once using ML. The results are presented in Tables \ref{simu_table_mult}, \ref{simu_table_add}, and \ref{simu_table_add_x}, where the boxes indicate the preferred model in terms of minimum bias (Mean) and maximum precision (minimum RMSE).

As in Table \ref{simu_table_mult}, for the Realized-ES-CAViaR-Mult model both MCMC and ML generate relatively accurate parameter estimates and VaR and ES forecasts, which proves the validity of both methods as discussed in Section \ref{beyesian_estimation_section}. Secondly, the results are in favor of the MCMC estimator: for 10 out of 11 parameters, as well as VaR forecasts, the MCMC approach generates lower bias and higher precision.

Regarding the Realized-ES-CAViaR-Add and Realized-ES-X-CAViaR-X models, as in Tables \ref{simu_table_add} and \ref{simu_table_add_x}, firstly we still observe that both MCMC and ML can generate relatively accurate VaR and ES forecasts. As discussed, although the true values of the $w_t$ equation $\gamma_0, \gamma_1, \gamma_2$ parameters could not be identified, for the other parameters both methods can produce good parameter estimates that are close to the true values. For both the Realized-ES-CAViaR-Add and Realized-ES-X-CAViaR-X models, overall MCMC is still preferred with more favorable bias and precision results in terms of the parameter estimation.

\begin{table}[!ht]
\begin{center}
\small
\caption{\label{simu_table_mult} \small Summary statistics for the two estimators of the Realized-ES-CAViaR-Mult model, with data simulated from Model (\ref{r_garch_simu}).}\tabcolsep=10pt
\begin{tabular}{lcccccccc} \hline
$n=1900$               &             & \multicolumn{2}{c}{MCMC}      &  \multicolumn{2}{c}{ML}   \\
Parameter              &True         &Mean           &  RMSE         &Mean           & RMSE    \\ \hline
$\beta_0$         &      -0.0465&   \fbox{-0.0631}	&\fbox{0.0790} &	-0.0655	&0.1201 \\
$\beta_1$         &    -0.2326& \fbox{-0.2505}&	\fbox{0.0930}	&-0.2496&	0.0979  \\
$\beta_2$          &     0.8500	&\fbox{0.8281}&	\fbox{0.0874}    &	0.8266&	0.1173  \\
$\xi$                  & 0.1000& \fbox{0.1835}&	\fbox{0.1689}	&0.1922	&0.1565\\
$\phi$              & 0.3377	& \fbox{0.2986}&	\fbox{0.1112} &0.2914	&0.1023\\
$\tau_{1}$               &0.0465	&  \fbox{0.0406} &	0.0159  & 0.0405&	\fbox{0.0158}  \\
$\tau_{2}$               &0.1082&   \fbox{0.0971}&	\fbox{0.0283}& 0.0969	&0.0284\\
$\sigma_{u}$           &0.3000&  \fbox{0.2801} &	\fbox{0.0203} &0.2797&	0.0208  \\
$\gamma_0$          &    -1.9264&  -2.1126&	\fbox{0.3467} &\fbox{-2.0640}&	0.5676   \\
$Q_{n+1}$     &    -1.2427	& \fbox{-1.2411}&	\fbox{0.0720}	&-1.2398&	0.0769    \\
$\text{ES}_{n+1}$     &    -1.4237&  -1.4018&	\fbox{0.0891}  & \fbox{-1.4066} &	0.0929    \\
\hline
\end{tabular}
\end{center}
\emph{Note}:\small A box indicates the favored estimators, based on mean and RMSE.
\end{table}

\begin{table}[!ht]
\begin{center}
\small
\caption{\label{simu_table_add} \small Summary statistics for the two estimators of the Realized-ES-CAViaR-Add model, with data simulated from Model (\ref{r_garch_simu}).}\tabcolsep=10pt
\begin{tabular}{lcccccccc} \hline
$n=1900$               &             & \multicolumn{2}{c}{MCMC}      &  \multicolumn{2}{c}{ML}   \\
Parameter              &True         &Mean           &  RMSE         &Mean           & RMSE    \\ \hline
$\beta_0$         &    -0.0465 & \fbox{-0.0611} &	 \fbox{0.0967} &	-0.2048 &	 0.5231 	   \\
$\beta_1$         &      -0.2326&   \fbox{-0.2471} &	 \fbox{0.0947}   & 	-0.2602 &	 0.1231   \\
$\beta_2$          &      0.8500 &	\fbox{0.8312} &	 \fbox{0.0978}   &	 0.7127 &	 0.3967    \\
$\xi$                  &  0.1000 &	 \fbox{0.1728} &	 \fbox{0.1424}    &	 0.1870 &	 0.5009  \\
$\phi$              &  0.3377 	 &    \fbox{0.3039} &	 \fbox{0.0943} 	& 0.2919 &	 0.3314   \\
$\tau_{1}$               &  0.0465 &  0.0404 &	 \fbox{0.0159} &	 \fbox{0.0442} &	 0.0438 	  \\
$\tau_{2}$               &    0.1082 	&  0.0966 &	 \fbox{0.0286}  & 	 \fbox{0.1047} & 	 0.0758  \\
$\sigma_{u}$           &  0.3000 	&  0.2801 &	 \fbox{0.0203}  &	 \fbox{0.2831} &	 0.0330  \\
$Q_{n+1}$     &       -1.2427 &   -1.2402 	& \fbox{0.0738}   &	\fbox{-1.2415} & 	 0.0885  \\
$\text{ES}_{n+1}$     &       -1.4237&  -1.4122 &	 \fbox{0.0852}     &	\fbox{-1.4284} &	0.1728   \\
\hline
\end{tabular}
\end{center}
\emph{Note}:\small  A box indicates the favored estimators, based on mean and RMSE.
\end{table}

\begin{table}[!ht]
\begin{center}
\small
\caption{\label{simu_table_add_x} \small Summary statistics for the two estimators of the Realized-ES-X-CAViaR-X model, with data simulated from Model (\ref{r_garch_simu}).}\tabcolsep=10pt
\begin{tabular}{lcccccccc} \hline
$n=1900$               &             & \multicolumn{2}{c}{MCMC}      &  \multicolumn{2}{c}{ML}   \\
Parameter              &True         &Mean           &  RMSE         &Mean           & RMSE    \\ \hline
$\beta_0$         &    -0.0465 &\fbox{-0.0660} &	\fbox{0.1283} 	  &	-0.0817 &	 0.2232    \\
$\beta_1$         &      -0.2326&    -0.2567 &	 \fbox{0.0968}   &	\fbox{-0.2488} &	 0.1047    \\
$\beta_2$          &      0.8500 &	 \fbox{0.8224} &	 \fbox{0.1221}  &	 0.8142 &	 0.1847     \\
$\xi$                  &  0.1000 &	 \fbox{0.1500} &	 \fbox{0.1386}     &	 0.1798 &	 0.1713  \\
$\phi$              &  0.3377 	 &    \fbox{0.3183} &	 \fbox{0.0935}  &	 0.2990 &	 0.1130  \\
$\tau_{1}$               &  0.0465 &  \fbox{0.0405} &	 0.0160  &	 0.0405 &	 \fbox{0.0158} 	  \\
$\tau_{2}$               &    0.1082 	&   0.0964 	 &0.0287   &	 \fbox{0.0967} &	 \fbox{0.0284}  \\
$\sigma_{u}$           &  0.3000 	&  \fbox{0.2802} &	 \fbox{0.0203}   &	 0.2796 &	 0.0208  \\
$Q_{n+1}$     &       -1.2427 &-1.2392 &	 \fbox{0.0728} & 	\fbox{-1.2394} &	 0.0788  \\
$\text{ES}_{n+1}$     &       -1.4237&   \fbox{-1.4128} &	 0.0904    &	-1.4096 &	\fbox{0.0883}  \\
\hline
\end{tabular}
\end{center}
\emph{Note}:\small  A box indicates the favored estimators, based on mean and RMSE.
\end{table}

{\centering
\section{\normalsize DETAILS OF REALIZED MEASURES}\label{realized_measure_section}
\par
}

\noindent
This section presents details of employed realized measures.

For day $t$, denoting the daily high, low, and closing prices as $H_{t}$, $L_{t}$, and $C_{t}$, the most commonly used daily log return is:
\begin{equation}\label{return_def}
r_t= \text{log}(C_t)-\text{log}(C_{t-1}), \nonumber
\end{equation}
where $r_t^2$ is the corresponding volatility estimator.

The high--low range (squared and scaled), proposed by Parkinson (1980), is a more efficient volatility estimator (see, for example, Feller, 1951):
\begin{equation}\label{range_def}
Ra_{t}^{2}=\frac {(\text{log}H_{t}-\text{log}L_{t})^2} {4\log2},
\end{equation}
where $4 \text{log}(2)$ makes $Ra_t$ approximately unbiased. The range allowing for overnight price jumps is discussed in Gerlach and Chen (2016):
\begin{equation} \label{range_o_def}
RaO_{t}= \text{log} \big( \text{max}(H_{t},C_{t-1})\big)-\text{log} \big( \text{min}(L_{t}, C_{t-1}),
\end{equation}
where again the volatility estimator squares $RaO_t$, then divides by $4 \text{log}(2)$.

Given each day $t$ is divided into $N$ equally sized intervals of length $\Delta$, subscripted by $\Theta= {0, 1, 2, ... , N}$, then several
high frequency volatility measures can be calculated. For day $t$, denote the $i$-$th$ interval closing price as $P_{t-1+i \triangle}$, and the interval high and low prices as
$H_{t,i}=\text{sup}_{(i-1) \triangle<j< i \triangle}P_{t-1+j}$ and $L_{t,i}=\text{inf}_{(i-1) \triangle<j<i \triangle}{P_{t-1+j}}$. Then RV is proposed by Andersen and Bollerslev (1998) as:
\begin{equation}\label{rv_def2}
RV_{t}^{\triangle}=\sum_{i=1}^{N} [log(P_{t-1+i \triangle})-log(P_{t-1+(i-1)\triangle})]^{2}.
\end{equation}

Martens and van Dijk (2007) and Christensen and Podolskij (2007) develop the Realized Range, which sums the squared intra-period
ranges:
\begin{equation}\label{rrv_def}
RR_{t}^{\triangle}= \frac {\sum_{i=1}^{N}(\text{log}H_{t,i}-\text{log}L_{t,i})^2}{4\log2}.
\end{equation}

Through theoretical derivation and simulation, Martens and van Dijk (2007) show that RR is a competitive, and sometimes more efficient,
volatility estimator than RV under some micro-structure conditions and levels. Gerlach and Wang (2016) confirm that RR can provide extra
efficiency in empirical tail risk forecasting, when employed as the measurement equation variable in an Realized-GARCH model.
To further reduce the effect
of microstructure noise, Martens and van Dijk (2007) present a scaling process, as follows:
\begin{eqnarray}\label{rv_scale}
ScRV_{t}^{\triangle}= \frac {\sum_{l=1}^{q}RV_{t-l}}{\sum_{l=1}^{q}RV_{t-l}^{\triangle}}RV_{t}^{\triangle},
\end{eqnarray}
\begin{eqnarray}\label{rrv_scale}
ScRR_{t}^{\triangle}= \frac {\sum_{l=1}^{q}RR_{t-l}}{\sum_{l=1}^{q}RR_{t-l}^{\triangle}}RR_{t}^{\triangle},
\end{eqnarray}
\noindent
where $RV_{t}$ and $RR_{t}$ represent the daily squared return and squared range on day $t$, respectively. This scaling process is
inspired by the fact that the daily squared return and range are less affected by micro-structure noise than their high frequency
counterparts, thus can be used to scale and smooth RV and RR, creating less micro-structure sensitive measures.

Further, Zhanget al. (2005) propose a sub-sampling process, also to deal with micro-structure effects.
For day $t$, $N$ equally sized samples are grouped into $M$ non-overlapping subsets $\Theta^{(m)}$ with size $N/M=n_{k}$, which means:
\begin{equation}
\Theta= \bigcup_{m=1}^{M} \Theta^{(m)}, \; \text{where} \; \Theta^{(k)}  \cap \Theta^{(l)} = \emptyset, \;
\text{when}  \;  k \neq l.  \nonumber
\end{equation}
Then sub-sampling will be implemented on the subsets $\Theta^{i}$ with $n_{k}$ interval:
\begin{equation}
\Theta^{i}= {i, i+n_k,...,i+n_k(M-2), i+n_k(M-1)}, \; \text{where} \;  i= {0,1,2...,n_k-1}.  \nonumber
\end{equation}

Representing the log closing price at the $i$-$th$ interval of day $t$ as $C_{t,i}=P_{t-1+i\triangle}$, the RV with the subsets
$\Theta^{i}$ is:
\begin{equation}
RV_{i}= \sum_{m=1}^{M} (C_{t,i+n_{k}m}-C_{t,i+n_{k}(m-1)})^{2}; \; \text{where} \; i= {0,1,2...,n_k-1}.  \nonumber
\end{equation}

We have the $T/M$ RV with $T/N$ sub-sampling as (supposing there are $T$ minutes per trading day):

\begin{equation}
RV_{T/M,T/N}= \frac{\sum_{i=0}^{n_k-1} RV_{i} } {n_k},
\end{equation}

then, denoting the high and low prices during the interval $i+n_{k}(m-1)$ and $i+n_{k}m$ as
$H_{t,i}=\text{sup}_{(i+n_{k}(m-1))\triangle<j<(i+n_{k}m) \triangle}P_{t-1+j}$ and
$L_{t,i}=\text{inf}_{(i+n_{k}(m-1))\triangle<j<(i+n_{k}m) \triangle}{P_{t-1+j}}$ respectively, we propose the $T/M$ RR with $T/N$ sub-sampling
as:
\begin{equation}
RR_{i}= \sum_{m=1}^{M} (H_{t,i}-L_{t,i})^2; \; \text{where} \; i= {0,1,2...,n_k-1},
\end{equation}
\begin{equation}
RR_{T/M,T/N}= \frac{\sum_{i=0}^{n_k-1} RR_{i} } {4 \text{log}2 n_k},
\end{equation}

For example, the 5 mins RV and RR with 1 min subsampling can be calculated as below respectively:
\begin{eqnarray}  \nonumber
&&RV_{5,1,0}=(\text{log} C_{t5}-\text{log} C_{t0})^2+(\text{log} C_{t10}-\text{log} C_{t5})^2+... \\  \nonumber
&&RV_{5,1,1}=(\text{log}C_{t6}-\text{log} C_{t1})^2+(\text{log} C_{t11}-\text{log} C_{t6})^2+... \\ \nonumber
&&RV_{5,1}=\frac{\sum_{i=0}^{4}RV_{5,1,i}} {5},\nonumber
\end{eqnarray}
\begin{eqnarray}  \nonumber
&&RR_{5,1,0}=(\text{log}H_{t0\leq t \leq t5}-\text{log} L_{t0\leq t \leq  t5})^2+(\text{log}H_{t5\leq t \leq t10}-\text{log}
L_{t5\leq t \leq  t10})^2+... \\ \nonumber
&&RR_{5,1,1}=(\text{log} H_{t1\leq t \leq t6}-\text{log} L_{t1\leq t \leq  t6})^2+(\text{log}H_{t6\leq t \leq t11}-\text{log}
L_{t6\leq t \leq  t11})^2+... \\ \nonumber
&&RR_{5,1}=\frac{\sum_{i=0}^{4}RR_{5,1,i}} {4 \text{log} (2)5}. \nonumber
\end{eqnarray}

Only intra-day return and range on the 5 minute frequency, additionally with 1 minute sub-sampling when employed, are considered in this paper. The properties of the sub-sampled RR, compared to those of other
realized measures, are assessed via simulation under three scenarios in Gerlach and Wang (2020b).

\end{appendices}

\end{document}